\documentclass[10pt,twocolumn]{IEEEtran}
\usepackage{mathrsfs}
\usepackage{color}
\usepackage{epsfig}
\usepackage{amsmath}
\usepackage{cite}
\usepackage{amsfonts, amssymb}
\usepackage{multirow}
\usepackage{array}
\newtheorem{theorem}{Theorem}

\newtheorem{lemma}{Lemma}

\newtheorem{assumption}{Assumption}
\newtheorem{remark}{Remark}

\def\BibTeX{{\rm B\kern-.05em{\sc i\kern-.025em b}\kern-.08em
T\kern-.1667em\lower.7ex\hbox{E}\kern-.125emX}}

\allowdisplaybreaks

\begin{document}
\title{Event-Triggered Resilient Filtering for 2-D Systems with Asynchronous-Delay: Handling Binary Encoding Decoding with Probabilistic Bit Flips
\thanks{This paragraph of the first footnote will contain the date on which you submitted your paper for review. This paper was produced by the IEEE Publication Technology Group. This work was supported by the National Natural Science Foundation of China (No. 62250056), the Natural Science Foundation of Shandong Province (No.ZR2021MF069, ZR2021JQ24).}
}
\author{Yu Chen and Wei Wang
\thanks{Y.~Chen and W.~Wang are with the School of Control Science and Engineering, Shandong University, Jinan 250061, P.~R.~China (Email: ychen$\_$2000@163.com, w.wang@sdu.edu.cn).}
}
\maketitle

\begin{abstract}
    In this paper, the event-triggered resilient filtering problem is investigated for a class of two-dimensional systems with asynchronous-delay under binary encoding-decoding schemes with probabilistic bit flips. To reduce unnecessary communications and computations in complex network systems, alleviate network energy consumption, and optimize the use of network resources, a new event-triggered mechanism is proposed, which focuses on broadcasting necessary measurement information to update innovation only when the event generator function is satisfied. A binary encoding-decoding scheme is used in the communication process to quantify the measurement information into a bit stream, transmit it through a memoryless binary symmetric channel with a certain probability of bit flipping, and restore it at the receiver. In order to utilize the delayed decoded measurement information that a measurement reconstruction approach is proposed. Through generating space equivalence verification, it is found that the reconstructed delay-free decoded measurement sequence contains the same information as the original delayed decoded measurement sequence. In addition, resilient filter is utilized to accommodate possible estimation gain perturbations. Then, a recursive estimator framework is presented based on the reconstructed decoded measurement sequence. By means of the mathematical induction technique, the unbiased property of the proposed estimator is proved. The estimation gain is obtained by minimizing an upper bound on the filtering error covariance. Subsequently, through rigorous mathematical analysis, the monotonicity of filtering performance with respect to triggering parameters is discussed.
\end{abstract}


\section{Introduction}
    \IEEEPARstart{I}{n} recent years, two-dimensional (2-D) systems have stimulated an ever-increasing research interest due to their application prospects in different scenarios, including but not limited to image analysis \cite{Ming_TCB21}, environmental monitoring systems \cite{Acharya_MTA21}, chromatography detection \cite{Muhlen_JSC06}, and wave propagation \cite{Andrade_JAM18}. Unlike the well-known one-dimensional (1-D) systems, information propagation in 2-D systems evolves independently in both horizontal and vertical directions. 2-D systems are usually represented by two types of models, namely the Fornasini-Marchesini state space model \cite{Fornasini_MST78} and the Roesser state space model \cite{Givone_ITC72}. To date, stability/performance analysis and control/filtering synthesis issues of various 2-D systems have been extensively studied. In \cite{Liu_TAC23}, under unknown but bounded external source interference and measurement noise, the zonotope-based interval estimation of a 2-D systems based on the event-triggered mechanism has been studied. In \cite{Cheng_TCB22}, the co-design scheme of asynchronous $H_{\infty}$ control for a class of 2-D Markov jump systems under adaptive event-triggered mechanism has been studied, by constructing nonlinear matrix inequalities, the asymptotic mean square stability of closed-loop system has been guaranteed. In \cite{Zhu_TAC23}, the design problem of cost-guaranteed controller for linear repetitive process event-triggered with multiplicative noises have been solved by utilizing chi-square distribution, complementary square technique, and actual information of triggering instants. In \cite{Yang_SMC19}, the dissipative control and filtering of a class of 2-D switching systems have been studied. By using the switching quadratic Lyapunov function technique, the sufficient conditions for the asymptotic stability of the system under arbitrary switching signals and the exponential stability of the 2-D TSR-$\vartheta$-dissipation rate under restricted switching signals has been given.
	
    For communication strategies transmitted over the network, event-triggered mechanisms (ETMs), as an effective method to save energy and bandwidth resources, are often used to control signal transmission from sensors to filter/controller. Compared with traditional time-triggered mechanism \cite{Geng_IF16,Zou_TAC19,Wang_SMC23} where the communication interval is known a priori, ETMs is an aperiodic communication protocol in which the transmission instant is determined by a defined event generator. Intuitively speaking, system components (such as sensors or actuators) are authorized to access the network to send signals only when they meet the triggering conditions. Otherwise, the signal will be held by the zero-order holder (ZOH) to maintain the signal value at the last triggering instant. So far, a large number of results have been reported on filter design issues under ETMs \cite{Zhao_TNSE22,Wang_TCB20,Chen_TNNLS23,Liu_TII15}. However, how to use event-triggered mechanism to achieve effective transmission and utilization of measurement information while considering multi-channel delay has not been fully studied. This is the main motivation of this article.
	
    With the rapid development of network communication technology and the widespread deployment of network devices, communication channels have brought multiple advantages to modern communication systems, including flexibility of architecture, ease of implementation, and low-cost maintenance. Nevertheless, due to limited communication capacity, transmitted signals are inevitably affected by network-induced phenomena. In order to improve the efficiency and reliability of data transmission, quantization-based encoding-decoding schemes (EDSs) have been widely used in the signal transmission process \cite{Azuma_ATO08,Brockett_TAC00,Leong_TSP13,Liu_ATO20,Leong_TSP13,Liu_TAC23}. Currently, because binary data is easy to implement and has transmission robustness, binary EDSs has become one of the widely used signal transmission strategies. Under binary EDSs, the transmission signal has been first encoded into a set of binary bit streams and transmitted through a memoryless binary symmetric channel. The final bit stream is restored to the transmission signal at the receiving end. During the transmission process, due to channel noise or malicious network attacks, bit flips may occur randomly with a certain crossover probability. If not handled properly, it will cause distortion of the transmission signal and lead to communication errors. Relevant results have appeared for 1-D systems \cite{Liu_TAC21,Wang_ATO23,Bernstein_ITIT66,Li_IS23,Hou_TNSE22,Shirazi_TSIP19,Zou_TAC23}, but there have been no relevant in-depth discussions on 2-D systems due to the difficulty of bidirectional evolution.
	
    In control systems that depend on network communication, delay is an inevitable problem. The causes of network delay is manifold, mainly including the distance of the transmission medium, network bandwidth, the processing and queuing of data packets, and hardware performance. In \cite{Liang_IS15}, the distributed $H_\infty$ state estimation problem of a class of 2-D systems with one-way stochastic delays have been studied. In \cite{Luo_IJGS15}, the problem of reliable estimation of 2-D discrete linear systems with one-way delays and incomplete observations have been studied. In \cite{Wang_SMC20}, the problem of robust filter design for a class of 2-D systems with norm-bounded parameter uncertainty and incomplete measurements (including random sensor delays and missing measurements) has been studied. The one-way random sensor delays have been presented in a unified form through the stochastic Kronecker function. Existing research on estimation problems of 2-D delays systems is all about one-way delays and is highly conservative. In \cite{Chen_CCC23}, the design scheme of a 2-D systems filter under two-way multi-step observation delays have been given for the first time. This paper will consider the more general state estimation of 2-D systems with asynchronous-delays.
	
    To tackle the aforementioned emerging issues, this paper investigates the estimation problem of 2-D systems with ETMs and asynchronous-delays under probabilistic bit flips binary EDSs. Some urgent challenges are outlined as follows: 1) How to propose a new approach to deal with the asynchronous-delays caused by the sensor transmission? 2) How to define an event-triggered shift sequence in view of the bidirectional signal propagations? 3) How to mathematically quantify the signal distortion during communication under binary EDSs? 4) How to design a recursive filter and present estimation gain? The main contributions of this paper can be highlighted as follows: {\it 1) In view of the asynchronous-delay, a decoded measurement reconstruction approach is proposed to achieve effective utilization of measurement information. 2) In order to reduce the communication burden of the network system, a new definition of triggering sequence and event generator function are proposed, based on which the occurrence of certain event is determined. 3) A new estimator framework is established for event-triggered filtering with asynchronous-delay under binary EDSs. 4) The impact of ETMs on filtering performance is discussed. In particular, a rigorous mathematical analysis of the monotonicity of filtering performance with respect to triggering parameters is performed.}

\section{Problem formulation}
    \subsection{System description}
        {\it Notations:} $\mathcal{E}\{\cdot\}$ represents the mathematical expectation. The scrip ``$\mathrm{T}$'' represents the transpose of vectors or matrices. $\mathbb{R}^n$ stands the $n$-dimensional Euclidean space. $\delta(i,j)$ is the Kronecker delta function with $\delta(i,j)$ being unity for $i=j$ but zero elsewhere. $\text{col}_{0\leq c \leq s}\{a_c\}$ represents a column vector with $a_0,a_1,\cdots,a_s$ as an element. $\text{diag}_{0\leq c \leq s}\{a_c\}$ stands for a diagonal matrix with $a_0,a_1,\cdots,a_s$ as the diagonal elements. $[0,k]$ represents the set $\{0,1,\cdots,k\}$. $\text{sym}(A)=A+A^\mathrm{T}$.

		Consider the following 2-D systems are described by FM-II model:
		\begin{align}\label{cw2:a1}
			x(i,j)&=f_1((i,j-1),x(i,j-1))\nonumber\\
			&\ +f_2((i-1,j),x(i-1,j))\nonumber\\
			&\ +B_1(i,j-1)w(i,j-1)\nonumber\\
			&\ +B_2(i-1,j)w(i-1,j)
		\end{align}
        where $i,j$ on a finite horizon $[0,\wp]$ with $\wp>0$ being a given integer, $x(i,j)\in \mathbb{R}^n$ is the system state vector and $w(i,j)\in\mathbb{R}^n$ is the process noise with zero-mean and covariance $Q(i,j)>0$. For $\iota\in[1,2]$, $B_{\iota}(i,j)$ are known shift-varying matrices with appropriate dimensions, the nonlinear functions $f_{\iota}((i,j),x(i,j))$ satisfying
		\begin{equation}\label{cw2:a2}
			\left\{
				\begin{aligned}
					&f_{\iota}((i,j),0)=0\\
					&\Vert f_{\iota}((i,j),x_1)-f_{\iota}((i,j),x_2)-A_{\iota}(i,j)(x_1-x_2)\Vert\\
					&\quad \leq a_{\iota}(i,j)\Vert x_1-x_2 \Vert \quad \forall x_1, x_2\in\mathbb{R}^n
				\end{aligned}
			\right.
		\end{equation}
		where $A_{\iota}(i,j)$ are known shift-varying matrices and $a_{\iota}(i,j)\geq 0$ are given scalars.
		
        In this study, the signal transmission between sensors and the remote estimator is conducted across multiple channels simultaneously, with the aim of enhancing the system fault tolerance. If a single channel fail or a signal be lost, other channels can serve as backups to ensure the complete transmission of the signal. However, given the restrictions on network bandwidth, this transmission method often leads to transmission delays. To accurately describe the process of signal transmission, the measurement model with asynchronous delays can be represented as:
        \begin{equation}\label{cw2:a3}
            \left\{
                \begin{aligned}
                    &y_{(0)}(i,j)=C_{(0)}(i,j)x(i-\imath_0,j-\jmath_0)+v_{(0)}(i,j)\\
                    &y_{(1)}(i,j)=C_{(1)}(i,j)x(i-\imath_1,j-\jmath_1)+v_{(1)}(i,j)\\
                    &\qquad\qquad\ \vdots\\
                    &y_{(N)}(i,j)=C_{(N)}(i,j)x(i-\imath_N,j-\jmath_N)+v_{(N)}(i,j)
                \end{aligned}
            \right.
        \end{equation}
        where $y_{(s)}(i,j)\in\mathbb{R}^{m}(s\in[0,N])$ is the $s$+1-th delayed measurement information and $v_{(s)}(i,j)\in\mathbb{R}^{m}$ is measurement noise with zero-mean and covariance $R_{(s)}(i,j)>0,$ $C_{(s)}(i,j)$ is known shift-varying matrices with appropriate dimensions.

\begin{figure}[!ht]  
			\centering
			\includegraphics [scale=0.5]{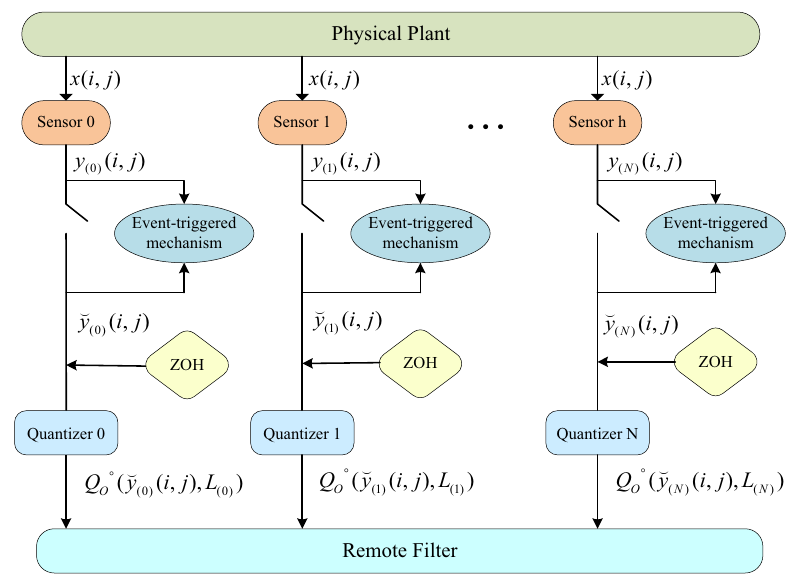}
			\caption{Framework of ETMs and EDSs for 2-D systems estimation.}
			\label{fig1}
		\end{figure}

        \begin{assumption}
            Without loss of generality, the delay $\imath_s$ and $\jmath_s$ are assumed to be in a increasing order: $0=\imath_0=\jmath_0<\imath_1\leq\jmath_1\cdots <\imath_N\leq\jmath_N$.
        \end{assumption}
		\begin{assumption}
            For $i,j\in[0,\wp], s\in[0,N]$, the initial state $x(i,0)$ and $x(0,j)$, random noises $w(i,j)$ and $v_{(s)}(i,j)$ are mutually uncorrelated with each other. The system \eqref{cw2:a1} initial condition satisfy the following statistical constraints:
			\begin{align*}
				&\mathcal{E}\{x(i,0)\}=x_u(i,0) \quad \mathcal{E}\{x(0,j)\}=x_u(0,j)\\
                &\mathcal{E}\{[x^\mathrm{T}(i,0)\ x^\mathrm{T}(0,j)]^\mathrm{T}\ [x^\mathrm{T}(r,0)\ x^\mathrm{T}(0,c)]\}\\
                &\ =diag\{\Xi(i,0)\delta(i,r),\Xi(0,j)\delta(j,c)\}
			\end{align*}
            where $x_u(i,0), x_u(0,j), \Xi(i,0)$, and $\Xi(0,j)$ are known shift-varying matrices with appropriate dimensions.
		\end{assumption}
        \begin{remark}
            In actual systems, due to factors such as equipment aging, working environment effects or cognitive limitations, it may lead to problems such as unknown model parameters or inaccurate descriptions, especially in the modeling of artificial systems and maneuverable targets. For example, In the dynamic evolution process of wave equation describing string vibration problem and heat equation describing water heating process, due to the existence of damped oscillation, elastic deformation and friction, a nonlinear model in which the displacement change of the underlying system is coupled with the system state will appear.
		\end{remark}
		\begin{remark}
            The framework of the estimation problem with the ETMs and EDSs for 2-D systems are presented in Fig.~\ref{fig1}. Sensors are only authorized to access the network and send measurement information when triggering conditions are met. Otherwise, the signal will be held by the ZOH to hold the measurement information at the last triggering instant. Then, the triggering measurement information is encoded into a set of binary bit streams and transmitted through a memoryless binary symmetric channel. Finally, the bit stream is restored to decoded measurement information at the receiver. All decoded measurement information is sent to the remote estimator for reconstruction.
		\end{remark}

	\subsection{Dynamic event-triggered mechanisms}
        Considering that periodic or continuous data transmission in the communication network, the shared network is occupied, resulting in a large number of redundant measurement outputs and causing network congestion. In addition, when there are only small fluctuations between adjacent measurement outputs, the filter does not need to consume resources again to update. Based on the above considerations, in order to reduce the data transmission frequency and improve network utilization efficiency, ETMs are used to determine whether to transmit the current measurement output to the remote filter.
		
		Define the event generator function as follows:
		\begin{align}\label{cw2:b1}
			\varphi_{(s)}^{\gamma}(i,j)\triangleq\varsigma_{(s)} h_{(s)}^{\gamma}(i,j)+\xi_{(s)}^{\gamma}(i,j)
		\end{align}
		where
		\begin{align*}
			h_{(s)}^{\gamma}(i,j)=&\sigma_{(s)}^{\gamma}(y_{(s)}^{\gamma}(i,j))^2-\rho_{(s)}^{\gamma}(e_{(s)}^{\gamma}(i,j))^2\nonumber\\
			e_{(s)}^{\gamma}(i,j)=&y_{(s)}^{\gamma}(i,j)-y_{(s)}^{\gamma}(i_t,j_t)
		\end{align*}
        for any $\gamma\in[1,m]$, $y_{(s)}^{\gamma}(i,j)$ and $y_{(s)}^{\gamma}(i_t,j_t)$ represents the $\gamma$-th component of the current measurement and the measurement received at the latest triggering instant $(i_t,j_t)$, respectively. $e_{(s)}^{\gamma}(i,j)$ represent the $\gamma$-th component of the dynamic event-triggered error $e_{(s)}(i,j)$. $\varsigma_{(s)},\rho_{(s)}^{\gamma},$ and $\sigma_{(s)}^{\gamma}$ represent triggering parameters. $\xi_{(s)}^{\gamma}(i,j)$ represent the $\gamma$-th component of the internal dynamic variable $\xi_{(s)}(i,j)$ governed by the following dynamics:
		\begin{align}\label{cw2:b2}
			\xi_{(s)}^{\gamma}(i,j)=&\alpha_{(s),1}\xi_{(s)}^{\gamma}(i,j-1)+\alpha_{(s),2}\xi_{(s)}^{\gamma}(i-1,j)\nonumber\\
			&+h_{(s)}^{\gamma}(i,j-1)+h_{(s)}^{\gamma}(i-1,j)
		\end{align}
        where $\alpha_{(s),1}$ and $\alpha_{(s),2}$ represent the pre-assigned scalars, and $\xi_{(s)}^{\gamma}(i,0)=\xi_{(s)}^{\gamma}(0,j)=\xi_{(s)}^0\geq0$ are given initial conditions of the internal dynamic variable $\xi_{(s)}^{\gamma}(i,j)$.

        It is worth noting that two indices are used in \eqref{cw2:b1}, and the sorting problem of generating index sequences will be explained below, prior to this, introduce a definition of partial order on integer pairs
		\begin{equation*}
			\left\{
				\begin{aligned}
						&(i_{t+1},j_{t+1})>(i_t,j_t),\quad i_{t+1}=i_t, j_{t+1}>j_t \\
                        &\qquad\qquad\qquad\qquad\quad\text{or}\  i_{t+1}>i_t,\  j_t\in[0,\wp],\\
						&(i_{t+1},j_{t+1})=(i_t,j_t),\quad i_{t+1}=i_t,j_{t+1}=j_t.			
				\end{aligned}
			\right.
		\end{equation*}
		Based on the definition of partial order mentioned above, the following triggering sequences are proposed:
		\begin{align*}
			(0,0)\leq(i_1,j_1)<(i_2,j_2)<\cdots<(i_t,j_t)<\cdots
		\end{align*}
		The forthcoming triggering instant $(i_{t+1},j_{t+1})$ is determined by
		\begin{align}\label{cw2:b3}
			(i_{t+1},j_{t+1})=inf&\big\{(i,j) | (i,j)>(i_t,j_t),\nonumber\\
			&\ \exists s, \gamma, s.t.\ \varphi_{(s)}^{\gamma}(i,j)\leq 0\big\}
		\end{align}
		\begin{remark}
            In view of the two-way propagation characteristics of the 2-D systems,This paper proposes a more general dynamic ETMs for 2-D systems, The event generator function $\varphi_{(s)}^{\iota}(i,j)$ has two indicators to represent the triggering instant $(i_t, j_t)$. Then, the trigger threshold is closely connected with system information and can be dynamically adjusted according to the operation of the system. When the measurement in the recent period fluctuates greatly, the internal dynamic variable $\xi_{(s)}^{\iota}(i,j)$ will be dynamically adjusted to reduce the trigger threshold to ensure that the current measurement satisfies the event trigger generator to implement trigger transmission. For the convenience of subsequent analysis, define $ \check{y}_{(s)}(i,j)=y_{(s)}(i_t,j_t),(i_t,j_t)\leq(i,j)<(i_{t+1},j_{t+1})$.
		\end{remark}
		\begin{remark}
            It should be noted that in the ETMs, the forthcoming triggering instant is not set independently in each channel. Instead, when there is a channel that meets the trigger execution conditions, all channels update the triggering instant and the current measurement is transmitted to the remote filter. This is different from the traditional ETMs. Considering the impact of delays on measurement, the ETMs adopted in this paper is conducive to the reconstruction of subsequent delayed measurement.
		\end{remark}
        \subsection{Encoding-decoding schemes}
            Consider using probabilistic bit flips EDSs to transmit the triggered measurement $\check{y}_{(s)}(i,j)$. More specifically, first utilize an encoder to preprocess the triggered measurement $\check{y}_{(s)}(i,j)$, i.e.,
			\begin{align*}
                \mathcal{Q}(\cdot):\check{y}_{(s)}^\gamma(i,j)\longrightarrow \mathcal{Q}(\check{y}_{(s)}^\gamma(i,j),L_{(s)})
			\end{align*}
            where $\check{y}_{(s)}^\gamma(i,j)$ represents the $\gamma$-th component of $\check{y}_{(s)}(i,j)$, $\gamma\in[1,m]$ . Let the value range of the triggered measurement $\check{y}_{(s)}^{\gamma}(i,j)$ at the horizon $(i,j)$ be $[-Z_{(s)},Z_{(s)}]$, among them, $Z_{(s)}\in\mathbb{R}>0$ is a known scalar determined by the range of the sensor $s$. By dividing $[-Z_{(s)},Z_{(s)}]$ into uniform length intervals $\Delta_{(s)}$, the triggered measurement $\check{y}_{(s)}^{\gamma}(i,j)$ is encoded as a binary bit string of length $L_{(s)}$,
			\begin{align*}
				\Delta_{(s)}=(2Z_{(s)})/(2^{L_{(s)}}-1)
			\end{align*}
		   when $c\Delta_{(s)} \leq \check{y}_{(s)}^{\gamma}(i,j)\leq (c+1)\Delta_{(s)}$,
            $\mathcal{Q}(\check{y}_{(s)}^{\gamma}(i,j),L_{(s)})=c\Delta_{(s)}$ or $\mathcal{Q}(\check{y}_{(s)}^{\gamma}(i,j),L_{(s)})=(c+1)\Delta_{(s)}$. The encoded output $\mathcal{Q}(\check{y}_{(s)}^{\gamma}(i,j),L_{(s)})$ is generated in the following probabilistic manner:
			\begin{equation}\label{cw2:c1}
				\left\{
					\begin{aligned}
                        &\text{Pr}\{\mathcal{Q}(\check{y}_{(s)}^{\gamma}(i,j),L_{(s)})=c\Delta_{(s)}\}=1-p_{(s)}^{\gamma}(i,j)\\
                        &\text{Pr}\{\mathcal{Q}(\check{y}_{(s)}^{\gamma}(i,j),L_{(s)})=(c+1)\Delta_{(s)}\}=p_{(s)}^{\gamma}(i,j)
					\end{aligned}
				\right.
			\end{equation}
            where $p_{(s)}^{\gamma}\triangleq (\check{y}_{(s)}^{\gamma}(i,j)-c\Delta_{(s)})/\Delta_{(s)}$ and $0\leq p_{(s)}^{\gamma}(i,j)\leq 1$.\\ Furthermore, the encoded output $\mathcal{Q}(\check{y}_{(s)}^{\gamma}(i,j),L_{(s)})$ can be represented in binary bits,
			\begin{align}\label{cw2:c2}
                \mathcal{Q}(\check{y}_{(s)}^{\gamma}(i,j),L_{(s)})=-Z_{(s)}+\sum_{\nu=1}^{L_{(s)}}b_{(s),\nu}^{\gamma}(i,j)2^{\nu-1}\Delta_{(s)}
			\end{align}
            where $\check{y}_{(s)}^{\gamma}(i,j)$ is encoded as a binary bit string $\{b_{(s),1}^{\gamma}(i,j),\\ b_{(s),2}^{\gamma}(i,j),\cdots,b_{(s),L_{(s)}}^{\gamma}(i,j)\}$ and $b_{(s),\nu}^{\gamma}(i,j)\in\{0,1\}$ is an encoded bit stream. Such flow is transmitted through the memoryless binary symmetric channel, and bit flipping may occur due to channel noise. Correspondingly, the received bit stream written is $\{\vec{b}_{(s),1}^{\gamma}(i,j), \vec{b}_{(s),2}^{\gamma}(i,j),\cdots,\vec{b}_{(s),L_{(s)}}^{\gamma}(i,j)\}$ and $\vec{b}_{(s),\nu}^{\gamma}(i,j)\in\{0,1\}$ where
			\begin{align*}
				\vec{b}_{(s),\nu}^{\gamma}(i,j)=&\varrho_{(s),\nu}(i,j)(1-b_{(s),\nu}^{\gamma}(i,j))\\
				&+(1-\varrho_{(s),\nu}(i,j))b_{(s),\nu}^{\gamma}(i,j)
			\end{align*}
			where $\varrho_{(s),\mu}(i,j)$ is a random variable obeying the Bernoulli distribution,
			\begin{align*}
				\varrho_{(s),\mu}(i,j)=\left\{
					\begin{aligned}
						&1,\ \text{the}\ \nu\text{-th bit is flipped}\\
						&0,\ \text{the}\ \nu\text{-th bit is not flipped}
					\end{aligned}
				\right.
			\end{align*}
            For the convenience of subsequent discussion, assuming that $\varrho_{(s),\nu}(i,j)$ is white and mutually uncorrelated, and have the same crossover probabilities. $\text{Pr}\{\varrho_{(s),\nu}(i,j)=1\}=\varrho_{(s)}$. Decode the received bit string and restore the triggered measurements according to the following rules:
			\begin{align}\label{cw2:c3}
                \mathcal{Q}^{\circ}(\check{y}_{(s)}^{\gamma}(i,j),L_{(s)})=-Z_{(s)}+\sum_{\nu=1}^{L_{(s)}}\vec{b}_{(s),\nu}^{\gamma}(i,j)2^{\nu-1}\Delta_{(s)}
			\end{align}
		
			In light of bit encoding rules, it can be seen that the encoded output can be rewritten as
			\begin{align}\label{cw2:c4}
				\mathcal{Q}(\check{y}_{(s)}^{\gamma}(i,j),L_{(s)})=\check{y}_{(s)}^{\gamma}(i,j)+q_{(s)}^{\gamma}(i,j)
			\end{align}
            where $q_{(s)}^{\gamma}(i,j)=\mathcal{Q}(\check{y}_{(s)}^{\gamma}(i,j),L_{(s)})-\check{y}_{(s)}^{\gamma}(i,j)$ represents the encoding truncation error. On the basis of \eqref{cw2:c1}, $q_{(s)}^{\gamma}(i,j)$ obeying the Bernoulli distribution taking values at $p_{(s)}^{\gamma}(i,j)\Delta_{(s)}$ or $(p_{(s)}^{\gamma}(i,j)-1)\Delta_{(s)}$, i.e.,
			\begin{equation*}
				\left\{
					\begin{aligned}
                        &\text{Pr}\{q_{(s)}^{\gamma}(i,j)=-p_{(s)}^{\gamma}(i,j)\Delta_{(s)}\}=1-p_{(s)}^{\gamma}(i,j)\\
                        &\text{Pr}\{q_{(s)}^{\gamma}(i,j)=(1-p_{(s)}^{\gamma}(i,j))\Delta_{(s)}\}=p_{(s)}^{\gamma}(i,j)
					\end{aligned}
				\right.
			\end{equation*}
			The proof that the encoding truncation error is unbiased and the variance is bounded will be given later.
			\begin{remark}
                Under the consideration of probabilistic bit flips EDSs, transmission errors (including quantization encoding errors and flipping bit errors) are inevitably generated during the transmission process. However, the quantization encoding error can be reduced by extending the length of the binary encoding stream, and the bit flipping error can be mitigated by utilizing more reliable communication channels and reducing the crossover probability.
			\end{remark}
\section{PRELIMINARIES}
    \subsection{The internal dynamic variable properties}
        Below, we provide a lemma to illustrate that the interval dynamic variable $\xi^{\iota}(i,j)\geq0$ is always hold.
		\begin{lemma}
            Let the parameters satisfies $\rho_{(s)}\geq \frac{1}{\alpha_{(s),1}}, \rho_{(s)}\geq \frac{1}{\alpha_{(s),2}},$ $ 0<\alpha_{(s),1},\alpha_{(s),2},\sigma_{(s)}^{\gamma}<1$ for any $\gamma\in[1,m]$. Then, we have $\xi_{(s)}^{\gamma}(i,j)\geq0$ for $i,j\in[0,\wp]$.
		\end{lemma}
		\begin{IEEEproof}
			1) For $(i,j)=(i_t,j_t)$,  retrospecting the event generator function, it is easy to obtain that
			\begin{align*}
                \varphi_{(s)}^{\gamma}(i,j)=&\varsigma_{(s)}\sigma_{(s)}^{\gamma}(y_{(s)}^{\gamma}(i_t,j_t))^2+\xi_{(s)}^{\gamma}(i_t,j_t)\\
				h_{(s)}^{\gamma}(i,j)=&\sigma_{(s)}^{\gamma}(y_{(s)}^{\gamma}(i_t,j_t))^2
			\end{align*}
            Due to the initial conditions of the interval dynamic variable and $h_{(s)}^{\gamma}(i,j)\geq0$, so $\xi_{(s)}^{\gamma}(i,j)=\xi_{(s)}^{\gamma}(i_t,j_t)\geq0$ is valid. Then, it indicates that for $(i_t,j_t)\leq (i,j)<(i_{t+1},j_{t+1})$, $\varphi_{(s)}^{\gamma}\\ (i,j)>0$, it is easy to know that
			\begin{align*}
				h_{(s)}^{\gamma}(i,j)>-\frac{1}{\varsigma_{(s)}}\xi_{(s)}^{\gamma}(i,j)
			\end{align*}
			Next, combining with \eqref{cw2:b2}, we conclude that
            \begin{small}
			\begin{align*}
				\xi_{(s)}^{\gamma}(i,j)>&(\alpha_{(s),1}-\frac{1}{\varsigma_{(s)}})\xi_{(s)}^{\gamma}(i,j-1)\\
				&+(\alpha_{(s),2}-\frac{1}{\varsigma_{(s)}})\xi_{(s)}^{\gamma}(i-1,j)
			\end{align*}
            \end{small}
		
			2) For $(i_t,j_t)\leq (i,j) <(i_{t+1},j_{t+1})$, according to the above analysis, we have
            \begin{footnotesize}
			\begin{align*}
				\xi_{(s)}^{\gamma}(i,j)&>(\alpha_{(s),1}-\frac{1}{\varrho_{(s)}})^2\xi_{(s)}^{\gamma}(i,j-2)\\
                &+2(\alpha_{(s),1}-\frac{1}{\varsigma_{(s)}})(\alpha_{(s),2}-\frac{1}{\varsigma_{(s)}})\xi_{(s)}^{\gamma}(i-1,j-1)\\
				&+(\alpha_{(s),2}-\frac{1}{\varsigma_{(s)}})^2\xi_{(s)}^{\gamma}(i-2,j)\\
				&>\cdots\\
				&>(\alpha_{(s),1}-\frac{1}{\varsigma_{(s)}})^{j_t}\xi_{(s)}^{\gamma}(i,j_t)\\
                &+\sum_{t=j_t+1}^{j-1}i(j-t)(\alpha_{(s),1}-\frac{1}{\varsigma_{(s)}})^{j-t}(\alpha_{(s),2}-\frac{1}{\varsigma_{(s)}})^i\xi_{(s)}^{\gamma}(i_t,t)\\
                &+\sum_{t=i_t+1}^{i-1}j(i-t)(\alpha_{(s),1}-\frac{1}{\varsigma_{(s)}})^{j}(\alpha_{(s),2}-\frac{1}{\varsigma_{(s)}})^{i-t}\xi_{(s)}^{\gamma}(t,j_t)\\
				&+(\alpha_{(s),2}-\frac{1}{\varsigma_{(s)}})^{i_t}\xi_{(s)}^{\gamma}(i_t,j)\\
				&\geq0.
			\end{align*}
            \end{footnotesize}
			To this end, by the induction, it is easy to get that $\xi_{(s)}^{\gamma}(i,j)\geq0$ for $i,j\in[0,\wp]$.
		\end{IEEEproof}

        Then, constrain the dynamic event-triggered error $\delta_{(s)}(i,j)$ by constructing an auxiliary variable $\check{\xi}_{(s)}^{\gamma}(i,j)$. For $(i_t,j_t)\leq (i,j) <(i_{t+1},j_{t+1})$, the auxiliary variable governed by the following dynamics:
        \begin{small}
		\begin{equation}\label{cw3:a1}
			\left\{
				\begin{aligned}
                    \check{\xi}_{(s)}^{\gamma}(i+1,j+1)&=\alpha_{(s),1}\check{\xi}_{(s)}^{\gamma}(i+1,j)+\alpha_{(s),2}\check{\xi}_{(s)}^{\gamma}(i,j+1)\\
					&+\sigma_{(s)}^{\iota}(y_{(s)}^{\gamma}(i+1,j))^2+\sigma_{(s)}^{\gamma}(y_{(s)}^{\gamma}(i,j+1))^2\\
					\check{\xi}_{(s)}^{\gamma}(i_t,j)&=\xi_{(s)}^{\gamma}(i_t,j), \check{\xi}_{(s)}^{\gamma}(i,j_t)=\xi_{(s)}^{\gamma}(i,j_t)
				\end{aligned}
			\right.
		\end{equation}
        \end{small}
        $\check{\xi}_{(s)}^{\gamma}(i,0)=\check{\xi}_{(s)}^{\gamma}(0,j)=\xi_{(s)}^0\geq0$ is given initial conditions of the auxiliary variable $\check{\xi}_{(s)}^{\gamma}(i,j)$. The following lemma will illustrate that $\check{\xi}_{(s)}^{\gamma}(i,j)\geq\xi_{(s)}^{\gamma}(i,j)$.
		\begin{lemma}\label{lem2}
            For $i,j\in[0,\wp]$, the interval dynamic variable $\xi_{(s)}^{\gamma}(i,j)$ is not greater than the auxiliary variable $\check{\xi}_{(s)}^{\gamma}(i,j)$.
		\end{lemma}
		\begin{IEEEproof}
			Prove this lemma via using mathematical induction, for $(i_t,j_t)\leq (i,j) <(i_{t+1},j_{t+1})$.
			
			 {\it Initial\ Step}: In light of \eqref{cw2:b2} and \eqref{cw3:a1}, it is easy to obtain
            \begin{footnotesize}
			\begin{align*}
				\xi_{(s)}^{\gamma}(i_t+1,j_t+1)&=\alpha_{(s),1}\xi^{\gamma}_{(s)}(i_t+1,j_t)+\alpha_{(s),2}\xi^{\gamma}_{(s)}(i_t,j_t+1)\nonumber\\
				&+h^{\gamma}_{(s)}(i_t+1,j_t)+h^{\gamma}_{(s)}(i_t,j_t+1)\\
				\check{\xi}_{(s)}(i_t+1,j_t+1)&=\alpha_{(s),1}\check{\xi}^{\gamma}_{(s)}(i_t+1,j_t)+\alpha_{(s),2}\check{\xi}^{\gamma}_{(s)}(i_t,j_t+1)\nonumber\\
				&+\sigma_{(s)}^{\gamma}(y_{(s)}^{\gamma}(i_t+1,j_t))^2+\sigma_{(s)}^{\gamma}(y_{(s)}^{\gamma}(i_t,j_t+1))^2
			\end{align*}
            \end{footnotesize}
            Due to $\check{\xi}_{(s)}^{\gamma}(i_t,j_t+1)=\xi_{(s)}^{\gamma}(i_t,j_t+1), \check{\xi}_{(s)}^{\gamma}(i_t+1,j_t)=\xi_{(s)}^{\gamma}(i_t\\ +1,j_t)$ and the definition of $h_{(s)}^{\gamma}(i,j)$. It is obvious that
			\begin{align*}
				\check{\xi}_{(s)}^{\gamma}(i_t+1,j_t+1)\geq\xi_{(s)}^{\gamma}(i_t+1,j_t+1)
			\end{align*}
		
            {\it Inductive\ Step}: Let $\check{\xi}_{(s)}^{\gamma}(i,j)\geq\xi_{(s)}^{\gamma}(i,j)$ is hold for $(i,j)\in\{(l,k)|l\in[i_t+1,i_{t+1}),k\in[j_t+1,j_{t+1}),l+k=L\}$ with $L\in[i_t+j_t+2,i_{t+1}+j_{t+1}-3)$, then ,for $(i,j)\in\{(l,k)|l\in[i_t+1,i_{t+1}),k\in[j_t+1,j_{t+1}),l+k=L+1\}$, we has
            \begin{small}
			\begin{align*}
				\xi_{(s)}^{\gamma}(i,j)=&\alpha_{(s),1}\xi_{(s)}^{\gamma}(i,j-1)+\alpha_{(s),2}\xi_{(s)}^{\gamma}(i-1,j)\nonumber\\
				&+h_{(s)}^{\gamma}(i,j-1)+h_{(s)}^{\gamma}(i-1,j)\\
                \check{\xi}_{(s)}^{\gamma}(i,j)=&\alpha_{(s),1}\check{\xi}_{(s)}^{\gamma}(i,j-1)+\alpha_{(s),2}\check{\xi}_{(s)}^{\gamma}(i-1,j)\nonumber\\
				&+\sigma_{(s)}^{\gamma}(y_{(s)}^{\gamma}(i,j-1))^2+\sigma_{(s)}^{\gamma}(y_{(s)}^{\gamma}(i-1,j))^2.
			\end{align*}
            \end{small}
            It can be known from the hypothesis that $\check{\xi}_{(s)}^{\gamma}(i,j)\geq\xi_{(s)}^{\gamma}(i,j)$. Therefor, along the same lines, we can conclude $\check{\xi}_{(s)}^{\gamma}(i,j)\geq\xi_{(s)}^{\gamma}(i,j)$ for $i,j\in[0,\wp]$.
		\end{IEEEproof}

        Based on the Lemma \ref{lem2}, a constrained upper bound for dynamic triggered error is given.
        \begin{small}
		\begin{align*}
            (e_{(s)}^{\gamma}(i,j))^2&\leq\frac{\sigma_{(s)}^{\gamma}}{\rho_{(s)}^{\gamma}}(y_{(s)}^{\gamma}(i,j))^2+\frac{1}{\varsigma_{(s)}\rho_{(s)}^{\gamma}}\xi_{(s)}^{\gamma}(i,j)\nonumber\\ &\leq\frac{\sigma_{(s)}^{\gamma}}{\rho_{(s)}^{\gamma}}(y_{(s)}^{\gamma}(i,j))^2+\frac{1}{\varsigma_{(s)}\rho_{(s)}^{\gamma}}\check{\xi}_{(s)}^{\gamma}(i,j)\nonumber\\
			&\triangleq(\bar{e}_{(s)}^{\gamma}(i,j))^2.
		\end{align*}
        \end{small}
		Thereby, the local upper bound of the dynamic triggering error covariance can be obtained
		\begin{align}\label{cw3:a2}
            &\mathbb{E}\{e_{(s)}(i,j)(e_{(s)}(i,j))^\mathrm{T}\}\nonumber\\
            &\quad\leq\text{diag}_{1\leq\gamma\leq m}\{(\bar{e}_{(s)}^{\gamma}(i,j))^2\}\triangleq\delta_{(s)}(i,j).
		\end{align}

	\subsection{Effects of binary encoding-decoding schemes}
		\begin{lemma}
            In the EDSs process, the encoding truncation error $q_{(s)}^{\gamma}(i,j)$ is of zero-mean and bounded variance, i.e.
			\begin{align*}
                &\mathcal{E}\{q_{(s)}^{\gamma}(i,j)\}=0,\\ &\mathcal{E}\{(q_{(s)}^{\gamma}(i,j))^2\}\leq \frac{1}{4}\Delta_{(s)}^2.
			\end{align*}
		\end{lemma}
		\begin{IEEEproof}
			 Retrospecting the probability distribution of encoding truncation error, it is easy to conclude that
			 \begin{align*}
			 	\mathcal{E}\{q_{(s)}^{\gamma}(i,j)\}=&-p_{(s)}^{\gamma}(i,j)\Delta_{(s)}(1-p_{(s)}^{\gamma}(i,j))\\
			 	&+(1-p_{(s)}^{\gamma}(i,j))\Delta_{(s)}p_{(s)}^{\gamma}(i,j)\\
			 	=&0.
			 \end{align*}
		 	Similarly, it is easy to know
		 	\begin{align*}
                &\text{Pr}\{(q_{(s)}^{\gamma}(i,j))^2=(p_{(s)}^{\gamma}(i,j))^2\Delta_{(s)}^2\}=1-p_{(s)}^{\gamma}(i,j)\\
                &\text{Pr}\{(q_{(s)}^{\gamma}(i,j))^2=(1-p_{(s)}^{\gamma}(i,j))^2\Delta_{(s)}^2\}=p_{(s)}^{\gamma}(i,j)
		 	\end{align*}
	 		which means that
	 		\begin{align*}
                \mathcal{E}\{(q_{(s)}^{\gamma}(i,j))^2\}=&(p_{(s)}^{\gamma}(i,j))^2\Delta_{(s)}^2(1-p_{(s)}^{\gamma}(i,j))\\
	 			&+(1-p_{(s)}^{\gamma}(i,j))^2\Delta_{(s)}^2p_{(s)}^{\gamma}(i,j)\\
	 			\leq& \frac{1}{4}\Delta_{(s)}^2.
	 		\end{align*}
 			This completes the proof.
		\end{IEEEproof}
		\begin{lemma}
            In the EDSs process, the encoding output $\mathcal{Q}(\check{y}_{(s)}^{\gamma}(i,j),L_{(s)})$ generated by the uniformly randomized quantizer is unbiased.
		\end{lemma}
		\begin{IEEEproof}
            According to \eqref{cw2:c4} and the truncation error $q_{(s)}^{\gamma}(i,\\ j)$ is of zero-mean, it is easy to know that $\mathcal{Q}(\check{y}_{(s)}^{\gamma}(i,j),L_{(s)})$ is unbiased.
		\end{IEEEproof}
	
		An important property of memoryless binary symmetric channel is given below.
		\begin{lemma}\label{lem3}
            Suppose that the encoded output $\mathcal{Q}(\check{y}_{(s)}^{\gamma}(i,j),\\L_{(s)})$ is transmitted through a memoryless binary symmetric channel with a crossover probability of $\varrho_{(s)}$. Then, the mean and variance of the decoded output $\mathcal{Q}^{\circ}(\check{y}_{(s)}^{\gamma}(i,j),L_{(s)})$ are
            \begin{align*}
                &\mathcal{E}\{\mathcal{Q}^{\circ}(\check{y}_{(s)}^{\gamma}(i,j),L_{(s)})\}=(1-2\varrho_{(s)})\mathcal{Q}(\check{y}_{(s)}^{\gamma}(i,j),L_{(s)})\\ &Var\{\mathcal{Q}^{\circ}(\check{y}_{(s)}^{\gamma}(i,j),L_{(s)})\}=\varrho_{(s)}(1-\varrho_{(s)})\frac{4Z_{(s)}^2(2^{2L_{(s)}}-1)}{3(2^{L_{(s)}}-1)^2}
			\end{align*}
		    where the expectation is about the random variable $\varrho_{(s),\nu}(i,j)$.
		\end{lemma}
        \begin{IEEEproof}
			According to \eqref{cw2:c3}, one has
            \begin{small}
			\begin{align*}
                &\mathcal{E}\{\mathcal{Q}^{\circ}(\check{y}_{(s)}^{\gamma}(i,j),L_{(s)})\}=-Z_{(s)}+\sum_{\nu=1}^{L_{(s)}}\mathcal{E}\{b^{\gamma}_{(s),\nu}(i,j)\}2^{\nu-1}\Delta_{(s)}\\
				& =-Z_{(s)}+\sum_{\nu=1}^{L_{(s)}}\Big(\varrho_{(s)}(1-b_{(s),\nu}^{\gamma}(i,j))\\
				&\quad+(1-\varrho_{(s)})b_{(s),\nu}^{\gamma}(i,j)\Big)2^{\nu-1}\Delta_{(s)}\\
                &=\mathcal{Q}(y_{(s)}^{\gamma}(i,j),L_{(s)})+\varrho_{(s)}\sum_{\nu=1}^{L_{(s)}}\Big(1-2b_{(s),\nu}^{\gamma}(i,j)\Big)2^{\nu-1}\Delta_{(s)}.
			\end{align*}
            \end{small}
			The last term of the above equation can be rewritten as
			\begin{small}
            \begin{align*}
				&\varrho_{(s)}\sum_{\nu=1}^{L_{(s)}}\Big(1-2b_{(s),\nu}^{\gamma}(i,j)\Big)2^{\nu-1}\Delta_{(s)}\\
                &\quad=\varrho_{(s)}\Big(\sum_{\nu=1}^{L_{(s)}}2^{\nu-1}\Delta_{(s)}-2\sum_{\nu=1}^{L_{(s)}}b_{(s),\nu}^{\gamma}(i,j)2^{\nu-1}\Delta_{(s)}\Big)\\
                &\quad=2\varrho_{(s)}\Big(Z_{(s)}-\sum_{\nu=1}^{L_{(s)}}b_{(s),\nu}^{\gamma}(i,j)2^{\nu-1}\Delta_{(s)}\Big)
			\end{align*}
            \end{small}
            In view of \eqref{cw2:c2}, $\mathcal{E}\{\mathcal{Q}^{\circ}(\check{y}_{(s)}^{\gamma}(i, j),L_{(s)})\}=(1-2\varrho_{(s)}) \mathcal{Q}(\check{y}_{(s)}^{\gamma}(i,\\ j),L_{(s)})$ can be verified. Moreover, the variance of $\mathcal{Q}^{\circ}(\check{y}_{(s)}^{\gamma}(i,\\ j),L_{(s)})$ is:
			\begin{small}
            \begin{align*}
                &Var\{\mathcal{Q}^{\circ}(\check{y}_{(s)}^{\gamma}(i,j),L_{(s)})\}=\mathcal{E}\Big\{\Big(-Z_{(s)}+\sum_{\nu=1}^{L_{(s)}}b_{(s),\nu}^{\gamma}(i,j)\\
				&\quad\times
                2^{\nu-1}\Delta_{(s)}\Big)^2\Big\}-\Big(\mathcal{E}\{\mathcal{Q}^{\circ}(y_{(s)}^{\gamma}(i,j),L_{(s)})\}\Big)^2\\ &=\mathcal{E}\Big\{\Big(\sum_{\nu=1}^{L_{(s)}}\big(b_{(s),\nu}^{\gamma}(i,j)-\mathcal{E}\{b_{(s),\nu}^{\gamma}(i,j)\}\big)2^{\nu-1}\Delta_{(s)}\\ &\quad+\mathcal{E}\{\mathcal{Q}^{\circ}(y_{(s)}^{\gamma}(i,j),L_{(s)})\}\Big)^2\Big\}+\Big(\mathcal{E}\big\{\mathcal{Q}^{\circ}(y_{(s)}^{\gamma}(i,j),L_{(s)})\big\}\Big)^2\\				&=\mathcal{E}\Bigl\{\Big(\sum_{\nu=1}^{L_{(s)}}\big(b_{(s),\nu}^{\gamma}(i,j)-\mathcal{E}\{b_{(s),\nu}^{\gamma}(i,j)\}\big)2^{\nu-1}\Delta_{(s)}\Big)^2\Bigr\}
			\end{align*}
            \end{small}
            It is worth noting that $b_{(s),\mu}^{\gamma}(i,j)(\mu\in[1,L_{(s)}])$ are mutually independent with $b_{(s),\mu}^{\gamma}(i,j)\in\{0,1\}$. Therefore, the above equation can be further expressed as
			\begin{small}
            \begin{align*}
                &\mathcal{E}\Bigl\{\Big(\sum_{\nu=1}^{L_{(s)}}\big(b_{(s),\nu}^{\gamma}(i,j)-\mathcal{E}\{b_{(s),\nu}^{\gamma}(i,j)\}\big)2^{\nu-1}\Delta_{(s)}\Big)^2\Bigr\}\\ &=\sum_{\nu=1}^{L_{(s)}}\Bigl(\mathcal{E}\big\{(b_{(s),\nu}^{\gamma}(i,j))^2\big\}-\big(\mathcal{E}\{b_{(s),\nu}^{\gamma}(i,j)\}\big)^2\Bigr)2^{2\nu-2}\Delta_{(s)}^2\\
				&=\varrho_{(s)}(1-\varrho_{(s)})\frac{4Z_{(s)}^2(2^{2L_{(s)}}-1)}{3(2^{L_{(s)}}-1)^2}
			\end{align*}
            \end{small}
			This completes the proof.
		\end{IEEEproof}

        Based on Lemma~\ref{lem3}, the decoded output $\mathcal{Q}^{\circ}(\check{y}_{(s)}^{\gamma}(i,j),L_{(s)})$ received through the memoryless binary symmetric channel can be restated as
		\begin{align}\label{cw3:b1}
			&\mathcal{Q}^{\circ}(\check{y}_{(s)}^{\gamma}(i,j),L_{(s)})\nonumber\\
			&\ =(1-2\varrho_{(s)})\mathcal{Q}(\check{y}_{(s)}^{\gamma}(i,j),L_{(s)})+\varkappa_{(s)}^{\gamma}(i,j)
		\end{align}
        where $\varkappa_{(s)}^{\gamma}(i,j)\in\mathbb{R}$ is random variable with $\mathcal{E}\{\varkappa_{(s)}^{\gamma}(i,j)\}=0$ and $Var\{\varkappa_{(s)}^{\gamma}(i,j)\}=Var\{\mathcal{Q}^{\circ}(\check{y}_{(s)}^{\gamma}(i,j),L_{(s)})\}\triangleq\beth_{(s)}^{\gamma}$.
	\subsection{Original decoded measurement description}
        Let $\mathcal{Q}_o^{\circ}(y_s(l,k),L_s)$ denotes the original decoded measurement received at the $(l,k)$ horizon of the 2-D systems \eqref{cw2:a1} and \eqref{cw2:c3}. For the sake of clarity, define the following sets
		\begin{align*}
			&\mathcal{T}_0^+=\{(l,k)|\imath_0\leq l< \imath_1,\jmath_1\leq k\leq j\},\nonumber \\
			&\mathcal{T}_0^\circ\ =\{(l,k)|\imath_0\leq l< \imath_1,d_0\leq k< \jmath_1\},\nonumber \\
			&\mathcal{T}_0^-\ =\{(l,k)|\imath_1\leq l\leq i,d_0\leq k< \jmath_1\},\nonumber\\
			&\qquad\qquad\qquad\qquad\qquad \vdots \nonumber\\
			&\mathcal{T}_{N-1}^+=\{(l,k)|\imath_{N-1}\leq l< \imath_{N},\jmath_{N}\leq k\leq j\},\nonumber \\
			&\mathcal{T}_{N-1}^\circ=\{(l,k)|\imath_{N-1}\leq l< \imath_{N},\jmath_{N-1}\leq k< \jmath_N\},\nonumber \\
			&\mathcal{T}_{N-1}^-=\{(l,k)|\imath_{N}\leq l\leq i,\jmath_{N-1}\leq k< \jmath_{N}\},\nonumber\\
			&\mathcal{T}_{N}^\circ\quad =\{(l,k)|\imath_{N}\leq l\leq i,\jmath_{N}\leq k\leq j\},\nonumber\\
			&\mathcal{T}_s=\mathcal{T}_s^+\cup \mathcal{T}_s^\circ\cup \mathcal{T}_s^-(s\in[0,N-1]),
            \mathcal{T}_N=\mathcal{T}_{N}^\circ.\nonumber
		\end{align*}
		The original decoded measurement can be describe as
		\begin{align}\label{cw3:c1}
			\mathcal{Q}_o^{\circ}(y_s(l,k),L_s)\triangleq
			\begin{bmatrix}
				\mathcal{Q}^{\circ}(y_{(0)}(l,k),L_{(0)})\\
				\mathcal{Q}^{\circ}(y_{(1)}(l,k),L_{(1)})\\
				\vdots\\
				\mathcal{Q}^{\circ}(y_{(s)}(l,k),L_{(s)})
			\end{bmatrix}
		\end{align}
        where $(i_t,j_t)\leq(i,j)<(i_{t+1},j_{t+1})$, $s\in\{0,1,\cdots,N\}$ for $(l,k)\in \{\mathcal{T}_0,\mathcal{T}_1,\cdots,\mathcal{T}_N\}$.
        \begin{remark}
            We will utilize the original decoded output sequence to estimate the system state. In light of \eqref{cw2:a3}, the original decoded measurement $\mathcal{Q}_o^{\circ}(y_s(l,k),L_s)$ contain asynchronous-delay, the filter developed in \cite{Muhlen_JSC06} and \cite{Givone_ITC72} cannot be used to solve the proposed estimation problem in this paper. Thus, it is necessary to develop a new approach to address asynchronous-delay and propose a novel recursive filter.
		\end{remark}
    \subsection{Reconstruction of the decoded measurement}
        In order to detailed investigate the estimation problem with measurement delays, we only consider the case of $i\geq \imath_N, j\geq \jmath_N$. Other cases can be handled in the same way. For brevity, denote $i_s=i-\imath_{N-s},j_s=j-\jmath_{N-s}$ and define the following sets
		\begin{align*}
			&\mathcal{S}_0^\circ=\{(l,k)|0\leq l\leq i_0,0\leq k\leq j_0\},\\
			&\mathcal{S}_1^-=\{(l,k)|0\leq l\leq i_0,j_0<k\leq j_1\},\\
			&\mathcal{S}_1^\circ=\{(l,k)|i_0<l\leq i_1,j_0< k\leq j_1\},\\
			&\mathcal{S}_1^+=\{(l,k)|i_0<l\leq i_1,0\leq k\leq j_0\},\\
			& \qquad\qquad\qquad\quad \vdots\\
			&\mathcal{S}_{N}^-=\{(l,k)|0\leq l\leq i_{N-1},j_{N-1}<k\leq j_{N}\},\\
			&\mathcal{S}_{N}^\circ=\{(l,k)|i_{N-1}<l\leq i_{N},j_{N-1}< k\leq j_{N}\},\\
			&\mathcal{S}_{N}^+=\{(l,k)|i_{N-1}<l\leq i_{N},0\leq k\leq j_{N-1}\},\\
            &\mathcal{S}_0=\mathcal{S}_0^\circ,\mathcal{S}_s=\mathcal{S}_s^-\cup \mathcal{S}_s^\circ \cup\mathcal{S}_s^+(s\in[1,N]),
		\end{align*}

		The original decoded measurements can be reconstructed as follows:
		\begin{align}\label{cw3:d1}
			\mathcal{Q}_r^{\circ}(y_s(l,k),L_s)\triangleq
			\begin{bmatrix}
				\mathcal{Q}^{\circ}(y_{(0)}(l,k),L_{(0)})\\
				\mathcal{Q}^{\circ}(y_{(1)}(l+\imath_1,k+\jmath_1),L_{(1)})\\
				\vdots\\
				\mathcal{Q}^{\circ}(y_{(s)}(l+\imath_s,k+\jmath_s),L_{(s)})
			\end{bmatrix}
		\end{align}
        where $(i_t,j_t)\leq(i,j)<(i_{t+1},j_{t+1})$, $s\in\{0,1,\cdots,N\}$ for $(l,k)\in \{\mathcal{S}_N,\mathcal{S}_{N-1},\cdots,\mathcal{S}_0\}$.
		
        In view of the definition of the original measurement and the original encoded measurement, it is easy to conclude that the corresponding reconstruction are $y_s(l,k)\triangleq\text{col}_{0\leq c \leq s}\{y_{(c)}(l+\imath_c,k+\jmath_c)\},\mathcal{Q}_r(y_s(l,k),L_s)\triangleq\text{col}_{0\leq c \leq s}\{\mathcal{Q}(y_{(c)}(l+\imath_c,k+\jmath_c),L_{(c)})\}.$
	
		It is easy to know that $y_s(l,k)$ satisfy
		\begin{align}\label{cw3:d2}
			y_s(l,k)=&C_s(l,k)x(l,k)+v_s(l,k)
		\end{align}
        where $C_s(l,k)=\text{col}_{0\leq c \leq s}\{C_{(c)}(l+\imath_c,k+\jmath_c)\}$ and $v_s(l,k)=\text{col}_{0\leq c \leq s}\{v_{(c)}(l+\imath_c,k+\jmath_c)\}$. $v_s(l,k)$ is zero-mean white noise with covariance matrices $R_s(l,k)=diag_{0\leq c\leq s}\big\{R_{(c)}(l+\imath_c,k+\jmath_c)\big\}$. Obviously, there is delay-free existed in \eqref{cw3:d2}. The following lemma will show that the reconstructed decoded measurement sequence and the original decoded measurement sequence have the same information.
		\begin{lemma}\label{lem6}
            The linear space spanned by the reconstructed measurement sequences and the linear space spanned by the original measurement sequences are equivalent, i.e.
			\begin{equation*}
				\begin{aligned}
                    \mathcal{L}\big\{\mathcal{Q}_r^{\circ}(y_s(l,k),L_s)|\ (l,k)\in\mathcal{S}_N\cup\mathcal{S}_{N-1}\cup\cdots\cup\mathcal{S}_0\big\}=\\ \mathcal{L}\big\{\mathcal{Q}_o^{\circ}(y_s(l,k),L_s)|\ (l,k)\in \mathcal{T}_0\cup \mathcal{T}_1\cup\cdots \cup \mathcal{T}_{N}\big\}
				\end{aligned}
			\end{equation*}
		\end{lemma}
        \begin{IEEEproof}
            The proof resembles that in \cite{Chen_CCC23}, due to space limitations, it is unmentioned.
		\end{IEEEproof}
		\begin{remark}
            The new decoded measurement sequence, $\{\mathcal{Q}_r^{\circ}\\ (y_s(l,k),L_s)|(l,k)\in \mathcal{S}_N\cup \mathcal{S}_{N-1}\cup \cdots \cup \mathcal{S}_{1}\}$ is named as reconstructed measurement sequence of $\{\mathcal{Q}_o^{\circ}(y_s(l,k),L_s)|(l,k)\in \mathcal{T}_0\cup\mathcal{T}_1\cup \cdots \cup \mathcal{T}_{N}\} $. The reconstructed decoded measurement sequence is all about the state information at the same horizon. And it is obvious from Lemma~\ref{lem6} that the reconstructed decoding measurement sequence contains the same information as the original decoding measurement sequence.
		\end{remark}

        The next section will give a filter design framework based on the reconstructed decoding measurement sequence. In order to facilitate analysis, for $(i_t,j_t)\leq(i,j)<(i_{t+1},j_{t+1})$, define $\check{y}_s(l,k)\triangleq y_s(l_t,k_t),
			\mathcal{Q}_r(\check{y}_s(l,k),L_s)\triangleq\mathcal{Q}_r(y_s(l_t,k_t),L_s),\\
			\mathcal{Q}_r^{\circ}(\check{y}_s(l,k),L_s)\triangleq\mathcal{Q}_r^{\circ}(y_s(l_t,k_t),L_s)$

\section{MAIN RESULTS}
	\subsection{Filter development}
			In this section, a $N$+1-step recursive filter is proposed as follows:
			
			\begin{subequations}\label{cw4:a1}
				$\mathbf{Step\  1}$: For $(l,k)\in \mathcal{S}_0$
				\begin{align}
					\hat{x}_p^1(l,k)=&f_1((l,k-1),\hat{x}_u^1(l,k-1))\nonumber\\
					&+f_2((l-1,k),\hat{x}_u^1(l-1,k))\label{cw4:a1a}\\
					\hat{x}_u^1(l,k)=&\hat{x}_p^1(l,k)+(K_1(l,k)+\alpha_1(l,k))\nonumber\\
					&\times\big(\mathcal{Q}_r^{\circ}(\check{y}_{N}(l,k),L_{N})-C_{N}(l,k)\hat{x}_p^1(l,k)\big)\label{cw4:a1b}
				\end{align}
			\end{subequations}
            where $\hat{x}_p^1(l,k)$ is the one-step prediction of the state $x(l, k)$, $\hat{x}_u^1(l,k)$ is the corresponding updated estimate with the initial
			conditions $\hat{x}_u^1(l,0)=x_u(l,0)$ and $\hat{x}_u^1(0,k)=x_u(0,k)$.
			
			\begin{subequations}\label{cw4:a2}
				$\mathbf{Step\  2}$: For $(l,k)\in \mathcal{S}_1$
				\begin{align}
					\hat{x}_p^2(l,k)=&f_1((l,k-1),\hat{x}_u^1(l,k-1))\nonumber\\
					&+f_2((l-1,k),\hat{x}_u^1(l-1,k))\label{cw4:a2a}\\
					\hat{x}_u^2(l,k)=&\hat{x}_p^2(l,k)+(K_2(l,k)+\alpha_2(l,k))\nonumber\\
                    &\times\big(\mathcal{Q}_r^{\circ}(\check{y}_{N-1}(l,k),L_{N-1})-C_{N-1}(l,k)\hat{x}_p^2(l,k)\big)\label{cw4:a2b}
				\end{align}
			\end{subequations}
            where $	\hat{x}_p^2(l,k)$ is the one-step prediction of the state $x(l,k)$, $\hat{x}_u^2(l,k)$ is the corresponding updated estimate with the initial conditions $\hat{x}_u^2(l,0)=x_u(l,0)(i_0+1\leq l \leq i_1),\hat{x}_u^2(l,j_0)=\hat{x}_u^1(l,j_0)(0 \leq l \leq i_0)$ and $\hat{x}_u^2(0,k)=x_u(0,k)(j_0+1 \leq k \leq j_1), \hat{x}_u^2(i_0,k)=\hat{x}_u^1(i_0,k)(0 \leq k \leq j_0)$.
			
			$\qquad\qquad\qquad\qquad\qquad \vdots$
			
			\begin{subequations}\label{cw4:a3}
				$\mathbf{Step\  N+1}$: For $(l,k)\in \mathcal{S}_{N}$
				\begin{align}
					\hat{x}_p^{N+1}(l,k)=&f_1((l,k-1),\hat{x}_u^{N+1}(l,k-1))\nonumber\\
					&+f_2((l-1,k),\hat{x}_u^{N+1}(l-1,k))\label{cw4:a3a}\\
					\hat{x}_u^{N+1}(l,k)=&\hat{x}_p^{N+1}(l,k)+(K_{N+1}(l,k)+\alpha_{N+1}(l,k))\nonumber\\
                    &\times\big(\mathcal{Q}_r^{\circ}(\check{y}_{0}(l,k),L_0)-C_{0}(l,k)\hat{x}_p^{N+1}(l,k)\big)\label{cw4:a3b}
				\end{align}
			\end{subequations}
            where $	\hat{x}_p^{N+1}(l,k)$ is the one-step prediction of the state $x(l,k)$, $\hat{x}_u^{N+1}(l,k)$ is the corresponding updated estimate with the initial conditions $\hat{x}_u^{N+1}(l,0)=x_u(l,0)(i_{N-1}+1\leq l \leq i_{N}), \hat{x}_u^{N+1}(l,j_{N-1})=\hat{x}_u^{N}(l,j_{N-1})(0 \leq l \leq i_{N-1})$ and $\hat{x}_u^{N+1}(0,k)=x_u(0,k) (j_{N-1}+1 \leq k \leq j_{N}), \hat{x}_u^{N+1}(i_{N-1},k)\\= \hat{x}_u^{N}(i_{N-1},k)(0 \leq k \leq j_{N-1})$.
			
            In \eqref{cw4:a1}-\eqref{cw4:a3}, $K_1(l,k), K_2(l,k), \cdots, K_{N+1}(l,k)$ are estimation gain parameters to be determined. $\alpha_1, \alpha_2, \cdots, \alpha_{N+1}$ are the estimation gain variation. $\alpha_{\tau}(l,k)=\sum_{\nu=1}^{r_{\tau}}\beta_{\tau,\nu} (i,j)H_{\tau,\nu}$, where
            $H_{\tau,\nu}\in\mathbb{R}^{n \times (N+1-\tau)m}$ is a set of known matrices and $\beta_{\tau,\mu}(i,j)$ is scalar-valued white noise with $\mathcal{E}\{\beta_{\tau,\nu}(i,j)\}=0$ and $Var\{\beta_{\tau,\nu}(i,j)\}=\check{\beta}_{\tau,\nu}$.
			\begin{remark}
                In control systems, accurate calculation of estimation gain is one of the key factors in ensuring system performance. Nevertheless, due to various reasons such as the non-ideal characteristics of electronic components, external interferences, and temperature changes, errors or uncertainties may arise in the calculation of estimation gain during the hardware implementation process. Such error or uncertainty can lead to a decrease in system performance or even cause system instability. Therefore, introducing the resilient characteristic in the estimation gain design is an effective strategy \cite{Acharya_MTA21}. Specifically, the introduction of the matrix $\alpha_{\tau}(l,k)$ to simulate gain perturbations allows for the consideration of possible changes and uncertainty in estimation gain during the design process, thereby enhancing the robustness of the control system.
			\end{remark}
			
			For the $N$+1-step recursive filter, for the sake of simplicity,
			we define
			\begin{align}
				\tilde{x}_p^{\tau}(l,k)&\triangleq x(l,k)-\hat{x}_p^{\tau}(l,k)\label{d1}\\
				\tilde{x}_u^{\tau}(l,k)&\triangleq x(l,k)-\hat{x}_u^{\tau}(l,k)\label{d2}\\
                P_p^{\tau}(l,k)&\triangleq \mathcal{E}\{\tilde{x}_p^{\tau}(l,k)(\tilde{x}_p^{\tau}(l,k))^\mathrm{T}\}\label{d3}\\
                P_u^{\tau}(l,k)&\triangleq \mathcal{E}\{\tilde{x}_u^{\tau}(l,k)(\tilde{x}_u^{\tau}(l,k))^\mathrm{T}\}\label{d4}\\
				X(l,k)&\triangleq \mathcal{E}\{x(l,k)x^\mathrm{T}(l,k)\}\label{d5}
			\end{align}
			Then, it follows from \eqref{cw2:a1}-\eqref{cw2:a3} and \eqref{cw4:a1}-\eqref{cw4:a3} that,
			
			\begin{subequations}\label{cw4:a4}
				$\mathbf{Step\  1}$: For $(l,k)\in \mathcal{S}_0$
				\begin{small}
                \begin{align}
					\tilde{x}_p^1(l,k)=&\tilde{f}_1((l,k-1),\tilde{x}_u^1(l,k-1))\nonumber\\
					&+\tilde{f}_2((l-1,k),\tilde{x}_u^1(l-1,k))+B_1(l,k-1)\nonumber\\
					&\times w(l,k-1)+B_2(l-1,k)w(l-1,k)\label{cw4:a4a}\\
                    \tilde{x}_u^1(l,k)=&\big[I-\big(K_1(l,k)+\alpha_1(l,k)\big)C_{N}(l,k)\big]\tilde{x}_p^1(l,k)\nonumber\\
					&-\big[K_1(l,k)+\alpha_1(l,k)\big]\big\{-2\varrho_{N} C_{N}(l,k)x(l,k)\nonumber\\
					&+(I-2\varrho_{N})\big[q_{N}(l,k)-e_{N}(l,k)\big]+\varkappa_{N}(l,k)\nonumber\\
					&+(I-2\varrho_{N})v_{N}(l,k)\big\}\label{cw4:a4b}
				\end{align}
                \end{small}
			\end{subequations}

            \begin{subequations}\label{cw4:a5}
				$\mathbf{Step\  2}$: For $(l,k)\in \mathcal{S}_1$
				\begin{small}
                \begin{align}
					\tilde{x}_p^2(l,k)=&\tilde{f}_1((l,k-1),\tilde{x}_u^2(l,k-1))\nonumber\\
					&+\tilde{f}_2((l-1,k),\tilde{x}_u^2(l-1,k))+B_1(l,k-1)\nonumber\\
					&\times w(l,k-1)+B_2(l-1,k)w(l-1,k)\label{cw4:a5a}\\
                    \tilde{x}_u^2(l,k)=&\big[I-\big(K_2(l,k)+\alpha_2(l,k)\big)C_{N-1}(l,k)\big]\tilde{x}_p^2(l,k)\nonumber\\
					&-\big[K_2(l,k)+\alpha_2(l,k)\big]\big\{-2\varrho_{N-1} C_{N-1}(l,k)x(l,k)\nonumber\\
					&+(I-2\varrho_{N-1})\big[q_{N-1}(l,k)-e_{N-1}(l,k)\big]+\varkappa_{N-1}(l,k)\nonumber\\
					&+(I-2\varrho_{N-1})v_{N-1}(l,k)\big\}\label{cw4:a5b}
				\end{align}
                \end{small}
			\end{subequations}
		
			$\qquad\qquad\qquad\qquad\qquad \vdots$
			
			\begin{subequations}\label{cw4:a6}
				$\mathbf{Step\  N+1}$: For $(l,k)\in \mathcal{S}_{N}$
				\begin{small}
                \begin{align}
					\tilde{x}_p^{N+1}(l,k)=&\tilde{f}_1((l,k-1),\tilde{x}_u^{N+1}(l,k-1))\nonumber\\
					&+\tilde{f}_2((l-1,k),\tilde{x}_u^{N+1}(l-1,k))+B_1(l,k-1)\nonumber\\
					&\times w(l,k-1)+B_2(l-1,k)w(l-1,k)\label{cw4:a6a}\\
                    \tilde{x}_u^{N+1}(l,k)=&\big[I-\big(K_{N+1}(l,k)+\alpha_{N+1}(l,k)\big)C_{0}(l,k)\Big]\tilde{x}_p^{N+1}(l,k)\nonumber\\
					&-\big[K_{N+1}(l,k)+\alpha_{N+1}(l,k)\big]\Big\{-2\varrho_0 C_{0}(l,k)x(l,k)\nonumber\\
					&+(I-2\varrho_0)\big[q_{0}(l,k)-e_{0}(l,k)\big]+\varkappa_{0}(l,k)\nonumber\\
					&+(I-2\varrho_0)v_{0}(l,k)\big\}\label{cw4:a6b}
				\end{align}
                \end{small}
			\end{subequations}
			for $\iota\in[1,2],\tau\in[1,N+1],s\in[0,N]$, where
			\begin{align*}
                &\tilde{f}_{\iota}((l,k),\tilde{x}_u^{\tau}(l,k))=f_{\iota}((l,k),x(l,k))-f_{\iota}((l,k),\hat{x}_u^{\tau}(l,k))\\
				&\varrho_s=\text{diag}_{0\leq c \leq s}\{\varrho_{(c)}\otimes I_m\}\\
				&e_s(l,k)=\text{col}_{0 \leq c \leq s}\{e_{(c)}(l+\imath_c,k+\jmath_c)\}\\
                &e_{(c)}(l+\imath_c,k+\jmath_c)=\text{col}_{1\leq \gamma \leq
                m}\{e_{(c)}^{\gamma}(l+\imath_c,k+\jmath_c)\}\\
				&q_s(l,k)=\text{col}_{0 \leq c \leq s}\{q_{(c)}(l+\imath_c,k+\jmath_c)\}\\
				&q_{(c)}(l+\imath_c,k+\jmath_c)=\text{col}_{1\leq \gamma \leq
                m}\{q_{(c)}^{\gamma}(l+\imath_c,k+\jmath_c)\}\\
				&\varkappa_s(l,k)=\text{col}_{0 \leq c \leq s}\{\varkappa_{(c)}(l+\imath_c,k+\jmath_c)\}\\
				&\varkappa_{(c)}(l+\imath_c,k+\jmath_c)=\text{col}_{1\leq \gamma \leq
                m}\{\varkappa_{(c)}^{\gamma}(l+\imath_c,k+\jmath_c)\}
			\end{align*}
            \begin{remark}
                It is worth noting that due to the existence of event triggering errors and encoding truncation errors, it is difficult to accurately calculate the error covariance dynamics in analytical form. Therefore, the target of this paper is to design an upper bound for the actual filtering error covariance, and to minimize this upper bound by appropriately designing estimation gains.
            \end{remark}	
    \subsection{Design of estimation gain parameters}
        In this subsection, we investigate estimation gains design problem for the resilient filter. Before presenting the main results, first, give the upper bound of the second-order moment of system state $x(l,k)$.
			\begin{lemma}
                Let $\check{a}_1,\check{a}_2>0$ are given scalars, the upper bound of the second-order moment of the state $x(l,k)$ is the solution of the following recursive evolution:
				\begin{align}\label{cw4:b1}
					\bar{X}(l,k)=&(1+\check{a}_1)\hat{X}_1(l,k-1)+(1+\check{a}_1^{-1})\hat{X}_2(l-1,k)\nonumber\\
					&+Q_1(l,k-1)+Q_2(l-1,k)
				\end{align}
                with initial condition $\bar{X}(l,0)=X(l,0)$ and $\bar{X}(0,k)= X(0,k\\)$, where $\hat{X}_{\iota}(l,k)\triangleq (1+\check{a}_2)a_{\iota}^2\text{tr}\{\bar{X}(l,k)\}I+(1+\check{a}_2^{-1})A_{\iota}(l,k)\\ \bar{X}(l,k)A_{\iota}^\mathrm{T}(l,k)$ and $Q_{\iota}(l,k)=B_{\iota}(l,k)Q(l,k)B_{\iota}^\mathrm{T}(l,k)$, $\iota=1,2$.
			\end{lemma}
			\begin{IEEEproof}
				According to \eqref{cw2:a1}-\eqref{cw2:a3} and the statistical property of $w(l,k)$
				\begin{align*}
					X(l,k)\leq & (1+\check{a}_1)\mathcal{E}\{f_1((l,k-1),x(l,k-1))\\
					&\times f_1^\mathrm{T}((l,k-1),x(l,k-1))\}+Q_1(l,k-1)\\
					&+(1+\check{a}_1^{-1}) \mathcal{E}\{f_2((l-1,k),x(l-1,k))\\
					&\times f_2^\mathrm{T}((l-1,k),x(l-1,k))\}+Q_2(l-1,k)
				\end{align*}
                Next, we will conduct a detailed analysis of $\mathcal{E}\{f_{\iota}((l,k),\\ x(l,k))f_{\iota}^\mathrm{T}((l,k),x(l,k))\}$,
				\begin{small}
                \begin{align*}
					&\mathcal{E}\{f_{\iota}((l,k),x(l,k))f_{\iota}^\mathrm{T}((l,k),x(l,k))\}\\
					&\quad=\mathcal{E}\Bigl\{\Big(f_{\iota}((l,k),x(l,k))-f_{\iota}((l,k),0)-A_{\iota}(l,k)x(l,k)\\
					&\quad\quad +A_{\iota}(l,k)x(l,k)\Big)\Big(f_{\iota}((l,k),x(l,k))-f_{\iota}((l,k),0)\\
					&\quad\quad -A_{\iota}(l,k)x(l,k)+A_{\iota}(l,k)x(l,k)\Big)^\mathrm{T}\Bigr\}\\
					&\quad \leq (1+\check{a}_2)\mathcal{E}\Bigl\{\Big(f_{\iota}((l,k),x(l,k))-f_{\iota}((l,k),0)\\
					&\quad\quad-A_{\iota}(l,k)x(l,k)\Big)\Big(f_{\iota}((l,k),x(l,k))-f_{\iota}((l,k),0)\\
                    &\quad\quad-A_{\iota}(l,k)x(l,k)\Big)\Bigr\}+(1+\check{a}_2^{-1})A_{\iota}(l,k)X(l,k)A_{\iota}^\mathrm{T}(l,k)\\
					&\quad\leq
                    (1+\check{a}_2)a_{\iota}^2\text{tr}\{X(l,k)\}I+(1+\check{a}_2^{-1})A_{\iota}(l,k)X(l,k)A_{\iota}^\mathrm{T}(l,k)
				\end{align*}
                \end{small}
				Based on the above derivation, it can be inferred that
				\begin{align}\label{cw4:b2}
					X(l,k)\leq& (1+\check{a}_1)X_1(l,k-1)+(1+\check{a}_2)X_2(l-1,k)\nonumber\\
					&+Q_1(l,k-1)+Q_2(l-1,k)
				\end{align}
                where $X_{\iota}(l,k)\triangleq (1+\check{a}_2)\text{tr}\{X(l,k)\}I+(1+\check{a}_2^{-1})A_{\iota}(l,k)\\X(l,k)A_{\iota}^\mathrm{T}(l,k)$.
                Then, using mathematical induction to prove that $\bar{X}(l,k)\geq X(l,k)$. To begin with, retrospecting the initial condition and \eqref{cw4:b2} that $\bar{X}(1,1)\geq X(1,1)$. Assume that $\bar{X}(l,k)\geq X(l,k)$ hold for $(l,k)\in\{(l_0,k_0)|l_0,k_0\in[1,\wp]; l_0\\ +k_0=\flat\}$ with $\flat\in[2,2\wp-1]$, then, for $(l,k)\in\{(l_0,k_0)|l_0,\\ k_0\in[1,\wp];l_0+k_0=\flat+1\}$, one has
				\begin{align*}
					&\bar{X}(l,k)-X(l,k)\\
					&\quad\geq (1+\check{a}_1)(\hat{X}_1(l,k-1)-X_1(l,k-1))\\
					&\quad\quad +(1+\check{a}_1^{-1})(\hat{X}_2(l-1,k)-X_2(l-1,k))\\
					&\quad\geq0
				\end{align*}
				This completes the proof.
			\end{IEEEproof}
            \begin{lemma}
                Let $\check{b}_1, \check{b}_2, \check{b}_3, \check{b}_4, \check{b}_5$, and $\check{b}_6$ are given positive scalar, the upper bound of the one-step prediction error covariance and the filtering error covariance are the solution of the following recursive evolution of the difference equations:
			
			\begin{subequations}\label{cw4:b3}
				$\mathbf{Step\  1}$: For $(l,k)\in \mathcal{S}_0$
				\begin{small}
                \begin{align}
					\Xi_p^1(l,k)=&(1+\check{b}_1)\Xi_{1p}^1(l,k-1)+(1+\check{b}_1^{-1})\Xi_{2p}^1(l-1,k)\nonumber\\
					&+Q_1(l,k-1)+Q_2(l-1,k)\label{cw4:b3a}\\
					\Xi_u^1(l,k)=&(1+\check{b}_3+\check{b}_4)\Big(\Xi_p^1(l,k)-K_1(l,k)C_{N}(l,k)\nonumber\\
					&\times \Xi_p^1(l,k)-\Xi_p^1(l,k)C_{\hbar}^\mathrm{T}(l,k)K_1^\mathrm{T}(l,k)\Big)\nonumber\\
					&+K_1(l,k)\check{\digamma}_{N}(\Xi_p^1(l,k))K_1^\mathrm{T}(l,k)\nonumber\\
                    &+\sum_{\nu=1}^{r_1}\check{\beta}_{1,\mu}H_{1,\mu}\check{\digamma}_{N}(\Xi_p^1(l,k))H_{1,\mu}^\mathrm{T}\label{cw4:b3b}
				\end{align}
                \end{small}
			\end{subequations}
			with the initial condition $\Xi_u^1(l,0)=P_u^1(l,0)$ and $\Xi_u^1(0,k)=P_u^1(0,k)$.
			
			\begin{subequations}\label{cw4:b4}
				$\mathbf{Step\  2}$: For $(l,k)\in \mathcal{S}_1$
                \begin{small}
				\begin{align}
					\Xi_p^2(l,k)=&(1+\check{b}_1)\Xi_{1p}^2(l,k-1)+(1+\check{b}_1^{-1})\Xi_{2p}^2(l-1,k)\nonumber\\
					&+Q_1(l,k-1)+Q_2(l-1,k)\label{cw4:b4a}\\
					\Xi_u^2(l,k)=&(1+\check{b}_3+\check{b}_4)\Big(\Xi_p^2(l,k)-K_2(l,k)C_{\hbar-1}(l,k)\nonumber\\
					&\times \Xi_p^2(l,k)-\Xi_p^2(l,k)C_{N-1}^\mathrm{T}(l,k)K_2^\mathrm{T}(l,k)\Big)\nonumber\\
					&+K_2(l,k)\check{\digamma}_{N-1}(\Xi_p^2(l,k))K_2^\mathrm{T}(l,k)\nonumber\\
                    &+\sum_{\nu=1}^{r_2}\check{\beta}_{2,\mu}H_{2,\mu}\check{\digamma}_{N-1}(\Xi_p^2(l,k))H_{2,\mu}^\mathrm{T}\label{cw4:b4b}
				\end{align}
                \end{small}
			\end{subequations}
            with the initial condition $\Xi_u^2(l,0)=P_u^2(l,0)(i_0+1\leq l \leq i_1)$, $\Xi_u^2(l,j_0)=\Xi_u^1(l,j_0)(0\leq l \leq i_0)$ and $\Xi_u^2(0,k)=P_u^2(0,k)(j_0+1\leq k \leq j_1)$, $\Xi_u^2(i_0,k)=\Xi_u^1(i_0,k)(0\leq k \leq j_0)$.
			
			$\qquad\qquad\qquad\qquad\qquad \vdots$
			
			\begin{subequations}\label{cw4:b5}
				$\mathbf{Step\  N+1}$: For $(l,k)\in \mathcal{S}_{N}$
				\begin{small}
                \begin{align}
                    \Xi_p^{N+1}(l,k)=&(1+\check{b}_1)\Xi_{1p}^{N+1}(l,k-1)+(1+\check{b}_1^{-1})\Xi_{2p}^{N+1}(l-1,k)\nonumber\\
					&+Q_1(l,k-1)+Q_2(l-1,k)\label{cw4:b5a}\\
					\Xi_u^{N+1}(l,k)=&(1+\check{b}_3+\check{b}_4)\Big(\Xi_p^{N+1}(l,k)-K_{N+1}(l,k)C_{0}(l,k)\nonumber\\
					&\times \Xi_p^{N+1}(l,k)-\Xi_p^{N+1}(l,k)C_{0}^\mathrm{T}(l,k)K_{N+1}^\mathrm{T}(l,k)\Big)\nonumber\\
					&+K_{N+1}(l,k)\check{\digamma}_{0}(\Xi_p^{N+1}(l,k))K_{N+1}^\mathrm{T}(l,k)\nonumber\\
                    &+\sum_{\nu=1}^{r_{N+1}}\check{\beta}_{{N+1},\mu}H_{{N+1},\mu}\check{\digamma}_{0}(\Xi_p^{N+1}(l,k))H_{{N+1},\mu}^\mathrm{T}\label{cw4:b5b}
				\end{align}
                \end{small}
			\end{subequations}
            with the initial condition $\Xi_u^{N+1}(l,0)=P_u^{N+1}(l,0)(i_{N-1}+1\leq l \leq i_{N})$, $\Xi_u^{N+1}(l, j_{N-1})=\Xi_u^{N}(l,j_{N-1})(0\leq l \leq i_{N-1})$ and $\Xi_u^{N+1}(0,k)=P_u^{N+1}(0,k)(j_{N-1}+1\leq k \leq j_{N})$, $\Xi_u^{N+1}(i_{N-1},k)=\Xi_u^{N}(i_{N-1},k)(0\leq k \leq j_{N-1})$.
			
			where
			\begin{small}
            \begin{align*}
				\Xi_{\iota p}^{\tau}(l,k)&=(1+\check{b}_2)a_{\iota}^2(l,k)\text{tr}\{\Xi_u^{\tau}(l,k)\}I\\
				& +(1+\check{b}_2^{-1})A_{\iota}(l,k)\Xi_u^{\tau}(l,k)A_{\iota}^\mathrm{T}(l,k)\\
                \check{\Theta}_{N+1-\tau}(\bar{X}(l,k))&=(1+\check{b}_3^{-1}+\check{b}_5)4\varrho_{N+1-\tau}C_{N+1-\tau}(l,k)\\
				&\times\bar{X}(l,k)C_{N+1-\tau}^\mathrm{T}(l,k)\varrho_{N+1-\tau}\\
				&+(I-2\varrho_{N+1-\tau})\vartheta_{N+1-\tau}(I-2\varrho_{N+1-\tau})\\
				&+(1+\check{b}_4^{-1}+\check{b}_5^{-1}+\check{b}_6)(I-2\varrho_{N+1-\tau})\\
				&\times \delta_{N+1-\tau}(l,k)(I-2\varrho_{N+1-\tau})+(1+\check{b}_6^{-1})\\
				&\times(I-2\varrho_{N+1-\tau})R_{N+1-\tau}(I-2\varrho_{N+1-\tau})\\
				&+\beth_{N+1-\tau}\\
                \check{\digamma}_{N+1-\tau}(\Xi_p^\tau(l,k))&=\check{\Theta}_{N+1-\tau}(\bar{X}(l,k))+(1+\check{b}_3+\check{b}_4)\\
				&\times C_{N+1-\tau}(l,k)\Xi_p^{\tau}(l,k)C_{N+1-\tau}^\mathrm{T}(l,k)
			\end{align*}
            \end{small}
		\end{lemma}
        \begin{IEEEproof}
			According to \eqref{d3} and \eqref{cw4:a4a}, it is easy to know that
			\begin{small}
            \begin{align}\label{cw4:b6}
                P_p^1(l,k)=&\mathcal{E}\big\{\tilde{f}_1\big((l,k-1),\tilde{x}_u^1(l,k-1)\big)\tilde{f}_1^\mathrm{T}\big((l,k-1),\nonumber\\
                &\tilde{x}_u^1(l,k-1)\big)\big\}+\mathcal{E}\big\{\tilde{f}_2\big((l-1,k),\tilde{x}_u^1(l-1,k)\big)\nonumber\\
				&\times \tilde{f}_2^\mathrm{T}\big((l-1,k),\tilde{x}_u^1(l-1,k)\big)\big\}+Q_1(l,k-1)\nonumber\\
                &+Q_2(l-1,k)+\text{sym}\Big(\mathcal{E}\big\{\tilde{f}_1\big((l,k-1),\tilde{x}_u^1(l,k-1)\big)\nonumber\\
				&\times\tilde{f}_2^\mathrm{T}\big((l-1,k),\tilde{x}_u^1(l-1,k)\big)\big\}\Big)
			\end{align}
            \end{small}
            Then, we will conduct a detailed analysis of $\mathcal{E}\big\{\tilde{f}_{\iota}\big((l,k),\\ \tilde{x}_u^1(l,k)\big)\tilde{f}_{\iota}^\mathrm{T}\big((l,k),\tilde{x}_u^1(l,k)\big)\big\}$,
			\begin{small}
            \begin{align*}
                &\mathcal{E}\big\{\tilde{f}_{\iota}\big((l,k),\tilde{x}_u^1(l,k)\big)\tilde{f}_{\iota}^\mathrm{T}\big((l,k),\tilde{x}_u^1(l,k)\big)\big\}\\
                &\quad =\mathcal{E}\Big\{\Big(\tilde{f}_{\iota}\big((l,k),\tilde{x}_u^1(l,k)\big)-(A_{\iota}(l,k)-A_{\iota}(l,k))\tilde{x}_u^1(l,k)\Big)\\
                &\quad\quad \times\Big(\tilde{f}_{\iota}\big((l,k),\tilde{x}_u^1(l,k)\big) -(A_{\iota}(l,k)-A_{\iota}(l,k))\tilde{x}_u^1(l,k)\Big)^\mathrm{T}\Big\}\\
                &\quad \leq(1+\check{b}_2)\mathcal{E}\Big\{\Big(\tilde{f}_{\iota}((l,k),\tilde{x}_u^1(l,k))-A_{\iota}(l,k)\tilde{x}_u^1(l,k)\Big)\\
                &\quad \quad \times \Big(\tilde{f}_{\iota}((l,k),\tilde{x}_u^1(l,k))-A_{\iota}(l,k)\tilde{x}_u^1(l,k)\Big)^\mathrm{T}\Big\}\\
				&\quad\quad +(1+\check{b}_2^{-1})A_{\iota}(l,k)P_u^1(l,k)A_{\iota}^\mathrm{T}(l,k)\\
				&\quad\leq (1+\check{b}_2)a_{\iota}^2(l,k)\text{tr}\{P_u^1(l,k)\}I\\
                &\quad\quad +(1+\check{b}_2^{-1})A_{\iota}(l,k)P_u^1(l,k)A_{\iota}^\mathrm{T}(l,k)\triangleq P_{\iota p}^1(l,k)
			\end{align*}
            \end{small}
            By using inequality $ab^\mathrm{T}+ba^\mathrm{T}\leq \check{b}_1aa^\mathrm{T}+\check{b}_1^{-1}bb^\mathrm{T}$, \eqref{cw4:b6} can be rewritten as
			\begin{align}\label{cw4:b7}
				P_p^1(l,k)\leq& (1+\check{b}_1)P_{1p}^1(l,k-1)+(1+\check{b}_1^{-1})P_{2p}^1(l-1,k)\nonumber\\
				&+Q_1(l,k-1)+Q_2(l-1,k)
			\end{align}
			According to \eqref{d4} and \eqref{cw4:a4b}, it is easy to know that
			\begin{small}
            \begin{align}\label{cw4:b8}
				P_u^1(l,k)=&\mathcal{E}\Big\{\big[I-\big(K_1(l,k)+\alpha_1(l,k)\big)C_{N}(l,k)\big]P_p^1(l,k)\nonumber\\
				&\times \big[I-\big(K_1(l,k)+\alpha_1(l,k)\big)C_{N}(l,k)\big]^\mathrm{T}\nonumber\\
                &+\big[K_1(l,k)+\alpha_1(l,k)\big]\Big\{4\varrho_{N} C_{N}(l,k)X(l,k)C_{N}^\mathrm{T}(l,k)\varrho_{N}\nonumber\\
				& +(I-2\varrho_{N})\big(\mathcal{E}\{q_N(l,k)q_N^\mathrm{T}(l,k)\}\nonumber\\
				&+\mathcal{E}\{e_{N}(l,k)e_{N}^\mathrm{T}(l,k)\}+R_{N}(l,k)\big)(I-2\varrho_{N})\nonumber\\
                &+\mathcal{E}\{\varkappa_{N}(l,k)\varkappa_{N}^\mathrm{T}(l,k)\}\Big\}[K_1(l,k)+\alpha_1(l,k)\big]^\mathrm{T}\nonumber\\
				&+\text{sym}\{\daleth_{11}+\daleth_{21}+\daleth_{31}+\daleth_{41}+\daleth_{51}\}\Big\}
			\end{align}
            \end{small}
			where
			\begin{small}
            \begin{align*}
				\daleth_{11}=&-\big[I-\big(K_1(l,k)+\alpha_1(l,k)\big)C_{N}(l,k)\big]\\
				&\times\mathcal{E}\Big\{\tilde{x}_p^1(l,k)\Big(-2\varrho_{N} C_{N}(l,k)x(l,k)+(I-2\varrho_{N})\\
				&\times\big[q_{N}(l,k)-e_{N}(l,k)+v_{N}(l,k)\big]+\varkappa_{N}(l,k)\Big)^\mathrm{T}\Big\}\\
				&\times\big[K_1(l,k)+\alpha_1(l,k)\big]^\mathrm{T}\\
                \daleth_{21}=&\big[K_1(l,k)+\alpha_1(l,k)\big]\mathcal{E}\Big\{-2\varrho_{N} C_{N}(l,k)x(l,k)\Big((I-2\varrho_{N})\\
				&\times \big[q_{N}(l,k)-e_{N}(l,k)+v_N(l,k)\big]+\varkappa_{N}(l,k)\Big)^\mathrm{T}\Big\}\\
				&\times\big[K_1(l,k)+\alpha_1(l,k)\big]^\mathrm{T}\\
                \daleth_{31}=&\big[K_1(l,k)+\alpha_1(l,k)\big]\mathcal{E}\Big\{(I-2\varrho_{N})q_{N}(l,k)\Big((I-2\varrho_{N})\\
				&\times \big[-e_{N}(l,k)+v_{N}(l,k)\big]+\varkappa_{N}(l,k)\Big)^\mathrm{T}\Big\}\\
				&\times\big[K_1(l,k)+\alpha_1(l,k)\big]^T\\
                \daleth_{41}=&\big[K_1(l,k)+\alpha_1(l,k)\big]\mathcal{E}\Big\{(I-2\varrho_{N})e_{N}(l,k)\Big((I-2\varrho_{N})\\
                &\times v_{N}(l,k)+\varkappa_{N}(l,k)\Big)^\mathrm{T}\Big\}\big[K_1(l,k)+\alpha_1(l,k)\big]^\mathrm{T}\\
                \daleth_{51}=&\big[K_1(l,k)+\alpha_1(l,k)\big]\mathcal{E}\Big\{(I-2\varrho_{N})v_{N}(l,k)\varkappa_{N}^\mathrm{T}(l,k)\Big\}\\
				&\times\big[K_1(l,k)+\alpha_1(l,k)\big]^\mathrm{T}
			\end{align*}
            \end{small}
            It is worth noting that $\tilde{x}_p^1(l,k)$ is independent of $q_{N}(l,k),\\ v_{N}(l,k)$, and $\varkappa_{N}(l,k)$. Can be derived that
			\begin{small}
            \begin{align*}
                \text{sym}\{\daleth_{11}\}\leq& (\check{b}_3+\check{b}_4)\big[I-\big(K_1(l,k)+\alpha_1(l,k)\big)C_{N}(l,k)\big]\\
				&\times P_p^1(l,k)\big[I-\big(K_1(l,k)+\alpha_1(l,k)\big)C_{N}(l,k)\big]^\mathrm{T}\\
				&+\check{b}_3^{-1}\big[K_1(l,k)+\alpha_1(l,k)\big]4\varrho_{N}C_{N}(l,k)\\
				&\times X(l,k)C_{N}^\mathrm{T}(l,k)\varrho_{N}\big[K_1(l,k)+\alpha_1(l,k)\big]^\mathrm{T}\\
				&+\check{b}_4^{-1}\big[K_1(l,k)+\alpha_1(l,k)\big](I-2\varrho_{N})\\
				&\times \mathcal{E}\{e_{N}(l,k)e_{N}^T(l,k)\}(I-2\varrho_{N})\big[K_1(l,k)+\alpha_1(l,k)\big]^\mathrm{T}.
			\end{align*}
            \end{small}
			Furthermore, $x(l,k)$ is independent of $q_{N}(l,k), v_{N}(l,k)$, and $\varkappa_{N}(l,k)$.
			\begin{align*}
				\text{sym}\{\daleth_{21}\}\leq&\check{b}_5\big[K_1(l,k)+\alpha_1(l,k)\big]4\varrho_{N}C_{N}(l,k)\\
				&\times X(l,k)C_{N}^\mathrm{T}(l,k)\varrho_{N}\big[K_1(l,k)+\alpha_1(l,k)\big]^\mathrm{T}\\
				&+\check{b}_5^{-1}\big[K_1(l,k)+\alpha_1(l,k)\big](I-2\varrho_{N})\mathcal{E}\{e_{N}(l,k)\\
				&\times e_{N}^\mathrm{T}(l,k)\}(I-2\varrho_{N})\big[K_1(l,k)+\alpha_1(l,k)\big]^\mathrm{T}
			\end{align*}
            It is known that $q_{N}(l,k), v_{N}(l,k)$, and $\varkappa_{N}(l,k)$ are mutually independent, thus, $\text{sym}\{\daleth_{31}+\daleth_{51}\}=0$,
			\begin{align*}
                \text{sym}\{\daleth_{41}\}\leq&\check{b}_6 \big[K_1(l,k)+\alpha_1(l,k)\big](I-2\varrho_{N})\mathcal{E}\{e_{N}(l,k)\\
				&\times e_{N}^\mathrm{T}(l,k)\}(I-2\varrho_{N})\big[K_1(l,k)+\alpha_1(l,k)\big]^\mathrm{T}\\
				&+\check{b}_6^{-1}\big[K_1(l,k)+\alpha_1(l,k)\big](1-2\varrho_{N})\\
				&\times R_{N}(l,k)(I-2\varrho_{N})\big[K_1(l,k)+\alpha_1(l,k)\big]^\mathrm{T}
			\end{align*}
			Substituting all cross terms into \eqref{cw4:b8} to obtain
			\begin{small}
            \begin{align*}
                P_u^1(l,k)\leq&\mathcal{E}\Big\{(1+\check{b}_3+\check{b}_4)\big[I-\big(K_1(l,k)+\alpha_1(l,k)\big)C_{N}(l,k)\big]\nonumber\\
				&\times P_p^1(l,k) \big[I-\big(K_1(l,k)+\alpha_1(l,k)\big)C_{N}(l,k)\big]^\mathrm{T}\nonumber\\
				&+(1+\check{b}_3^{-1}+\check{b}_5)\big[K_1(l,k)+\alpha_1(l,k)\big]4\varrho_{N}\\
				&\times C_{N}(l,k)X(l,k)C_{N}^\mathrm{T}(l,k)\varrho_{N}\big[K_1(l,k)+\alpha_1(l,k)\big]^\mathrm{T}\\
				&+\big[K_1(l,k)+\alpha_1(l,k)\big](I-2\varrho_{N})\mathcal{E}\{q_{N}(l,k)\\
				&\times q_{N}^\mathrm{T}(l,k)\}(I-2\varrho_{N})\big[K_1(l,k)+\alpha_1(l,k)\big]^\mathrm{T}\\
				&+(1+\check{b}_4^{-1}+\check{b}_5^{-1}+\check{b}_6)\big[K_1(l,k)+\alpha_1(l,k)\big](I-2\varrho_{N})\\
                &\times \mathcal{E}\{e_{N}(l,k)e_{N}^\mathrm{T}(l,k)\}(I-2\varrho_{N})\big[K_1(l,k)+\alpha_1(l,k)\big]^\mathrm{T}\\
				&+(1+\check{b}_6^{-1})\big[K_1(l,k)+\alpha_1(l,k)\big](I-2\varrho_{N})R_{N}(l,k)\\
				&\times (I-2\varrho_{N})\big[K_1(l,k)+\alpha_1(l,k)\big]^\mathrm{T}+\big[K_1(l,k)\\
				&+\alpha_1(l,k)\big] \mathcal{E}\{\varkappa_{N}(l,k)\varkappa_{N}^\mathrm{T}(l,k)\}\\
				&\times \big[K_1(l,k)+\alpha_1(l,k)\big]^\mathrm{T}\Big\}
			\end{align*}
            \end{small}
			The first term in the above equation can be decomposed to obtain,
			\begin{small}
            \begin{align*}
				&\mathcal{E}\Big\{\big[I-\big(K_1(l,k)+\alpha_1(l,k)\big)C_{N}(l,k)\big]P_p^1(l,k)\\
				&\quad \times[I-\big(K_1(l,k)+\alpha_1(l,k)\big)C_{N}(l,k)\big]^\mathrm{T}\Big\}\\
				&\ =P_p^1(l,k)-K_1(l,k)C_{N}(l,k)P_p^1(l,k)\\
                &\quad -P_p^1(l,k)C_{N}^\mathrm{T}(l,k)K_1^\mathrm{T}(l,k)+\mathcal{E}\Big\{\big(K_1(l,k)+\alpha_1(l,k)\big)\\
				&\quad \times C_{N}(l,k)P_p^1(l,k)C_{N}^\mathrm{T}(l,k)\big(K_1(l,k)+\alpha_1(l,k)\big)^\mathrm{T}\Big\}
			\end{align*}
            \end{small}
			and
			\begin{small}
            \begin{align*}
				&\mathcal{E}\Big\{\big(K_1(l,k)+\alpha_1(l,k)\big)C_{N}(l,k)P_p^1(l,k)\\
				&\qquad\times C_{N}^\mathrm{T}(l,k)\big(K_1(l,k)+\alpha_1(l,k)\big)^\mathrm{T}\Big\}\\
				&\quad=K_1(l,k)C_{N}(l,k)P_p^1(l,k)C_{N}^\mathrm{T}(l,k)K_1^\mathrm{T}(l,k)\\
                &\qquad+\sum_{\nu=1}^{r_1}\check{\beta}_{1,\nu}H_{1,\nu}C_{N}(l,k)P_p^1(l,k)C_{N}^\mathrm{T}(l,k)H_{1,\nu}^\mathrm{T}
			\end{align*}
            \end{small}
			In view of the ETMs,
			\begin{align*}
                \mathcal{E}\{e_{N}(l,k)e_{N}^\mathrm{T}(l,k)\}&\leq \text{diag}_{0\leq c \leq N}\{\delta_{(c)}(l+\imath_c,k+\jmath_c)\}\triangleq \delta_{N}(l,k).
			\end{align*}
			According to the statistical property of the truncation error,
			\begin{align*}
				&\mathcal{E}\{q_{N}(l,k)q_{N}^\mathrm{T}(l,k)\}\\
                &\ = \text{diag}_{0\leq c \leq N}\{\mathcal{E}\{q_{(c)}(l+\imath_c,k+\jmath_c)q_{(c)}^\mathrm{T}(l+\imath_c,k+\jmath_c)\}\}\\
				&\mathcal{E}\{q_{(c)}(l+\imath_c,k+\jmath_c)q_{(c)}^\mathrm{T}(l+\imath_c,k+\jmath_c)\}\\
                &\ = \text{diag}_{1 \leq \gamma \leq m}\{\mathcal{E}\{(q_{(c)}^\gamma(l+\imath_c,k+\jmath_c))^2\}\}\\
				&\ \leq \text{diag}_{1 \leq \gamma \leq m}\{\frac{1}{4}\Delta_{(c)}^2\}.
			\end{align*}
			In light of \eqref{cw3:b1}, it is easy to conclude that
			\begin{align*}
                &\mathcal{E}\{\varkappa_{N}(l,k)\varkappa_{N}^\mathrm{T}(l,k)\}\\
                &\ =\text{diag}_{0 \leq c \leq N}\{\mathcal{E}\{\varkappa_{(c)}(l+\imath_c,k+\jmath_c)\varkappa_{(c)}^\mathrm{T}(l+\imath_c,k+\jmath_c)\}\}\\
                &\mathcal{E}\{\varkappa_{(c)}(l+\imath_c,k+\jmath_c)\varkappa_{(c)}^\mathrm{T}(l+\imath_c,k+\jmath_c)\}\\
                &\ =\text{diag}_{1\leq \gamma \leq m}\{\mathcal{E}\{(\varkappa_{(c)}^\gamma(l+\imath_c,k+\jmath_c))^2\}\}\\
				&\ \leq \text{diag}_{1\leq \gamma \leq m}\{\beth_{(c)}^\gamma\}
			\end{align*}
			so, we can record briefly
			\begin{align*}
                \mathcal{E}\{q_{N}(l,k)q_{N}^\mathrm{T}(l,k)\}\leq \vartheta_{N},\quad \mathcal{E}\{\varkappa_{N}(l,k)\varkappa_{N}^\mathrm{T}(l,k)\}\leq \beth_{N}.
			\end{align*}
			Denote
			\begin{small}
            \begin{align*}
                \Theta_{N}(X(l,k))&=(1+\check{b}_3^{-1}+\check{b}_5)4\varrho_{N}C_{N}(l,k)X(l,k)C_{N}^\mathrm{T}(l,k)\varrho_{N}\\
				&+(I-2\varrho_{N})\vartheta_{N}(I-2\varrho_{N})+(1+\check{b}_4^{-1}+\check{b}_5^{-1}+\check{b}_6)\\
				&\times(I-2\varrho_{N})\delta_{N}(l,k)(I-2\varrho_{N})+(1+\check{b}_6^{-1})\\
				&\times (I-2\varrho_{N})R_{N}(l,k)(I-2\varrho_{N})+\beth_{N}\\
				\digamma_{N}(P_p^1(l,k))&=\Theta_{N}(X(l,k))+(1+\check{b}_3+\check{b}_4)\\
				&\times C_{N}(l,k)P_p^1(l,k)C_{N}^\mathrm{T}(l,k).
			\end{align*}
            \end{small}
			It is easy to know that
			\begin{align}\label{cw4:b9}
				P_u^1(l,k)\leq& (1+\check{b}_3+\check{b}_4)\big(P_p^1(l,k)-K_1(l,k)C_{N}(l,k)\nonumber\\
				&\times P_p^1(l,k)-P_p^1(l,k)C_{N}^\mathrm{T}(l,k)K_1^\mathrm{T}(l,k)\big)\nonumber\\
				&+K_1(l,k)\digamma_{N}(P_p^1(l,k))K_1^\mathrm{T}(l,k)\nonumber\\
				&+\sum_{\nu=1}^{r_1}\check{\beta}_{1,\nu}H_{1,\nu}\digamma_{N}(P_p^1(l,k))H_{1,\nu}^\mathrm{T}.
			\end{align}
            Then, by using mathematical induction to prove that $\Xi_p^1(l,k)\\ \geq P_p^1(l,k)$ and $\Xi_u^1(l,k)\geq P_u^1(l,k)$. According to the initial condition and \eqref{cw4:b9}, it is easy to know that $\Xi_p^1(1,1)\geq P_p^1(1,1)$, which imply $\Xi_u^1(1,1)\geq P_u^1(1,1)$. Next, assume $\Xi_u^1(l,k)\geq P_u^1(l,k)$ is valid for $(l,k)\in\{(l_0,k_0)|l_0,k_0\in[1,\wp]; l_0+k_0=\flat\}$ with $\flat\in[2,2\wp-1]$, then, for $(l,k)\in\{(l_0,k_0)|l_0,k_0\in[1,\wp];l_0+k_0=\flat+1\}$, one has
			\begin{small}
            \begin{align*}
				&\Xi_p^1(l,k)-P_p^1(l,k)\\
				&\ \geq (1+\check{b}_1)\big(\Xi_{1p}^1(l,k-1)-P_{1p}^1(l,k-1)\big)\\
				&\quad +(1+\check{b}_1^{-1})\big(\Xi_{2p}^1(l-1,k)-P_{2p}^1(l-1,k)\big)\\
				&\ \geq 0\\
				&\Xi_u^1(l,k)-P_u^1(l,k)\\
				&\ \geq (1+\check{b}_3+\check{b}_4)\big[I-\big(K_1(l,k)+\alpha_1(l,k)\big)C_{N}(l,k)\big]\\
                &\quad\times\big(\Xi_p^1(l,k)-P_p^1(l,k)\big)\big[I-\big(K_1(l,k)+\alpha_1(l,k)\big)C_{N}(l,k)\big]^\mathrm{T}\\
				&\quad +K_1(l,k)(\check{\Theta}_{N}(\bar{X}(l,k))-\Theta_{N}(X(l,k)))K_1^\mathrm{T}(l,k)\\
                &\quad+\sum_{\mu}^{r_1}\check{\beta}_{1,\mu}H_{1,\mu}\big(\check{\digamma}_{N}(\bar{\Xi}_p^1(l,k))-\digamma_{N}(\Xi_p^1(l,k))\big)H_{1,\mu}^\mathrm{T}\\
				&\ \geq 0
			\end{align*}
            \end{small}
			Similar methods can be used to prove other areas. This completes the proof of the lemma.
		\end{IEEEproof}

        The following theorem gives the estimation gain parameters by minimizing the local upper bound on the estimation error covariance.
		\begin{theorem}\label{the1}
            For the 2-D systems \eqref{cw2:b1} and \eqref{cw2:b3}, the recursive evolution of the upper bound of filtering error covariance and the estimation gains are given as follows,
			
			\begin{subequations}\label{t1-1}
				$\mathbf{Step\  1}$: For $(l,k)\in \mathcal{S}_0$
				\begin{small}
                \begin{align}
                    K_1(l,k)=&(1+\check{b}_3+\check{b}_4)\Xi_p^1(l,k)C_{N}^\mathrm{T}(l,k)\big(\check{\digamma}_{N}(\Xi_p^1(l,k))\big)^{-1}\label{t1-1a}\\
                    \Xi_u^1(l,k)=&(1+\check{b}_3+\check{b}_4)\Xi_p^1(l,k)-K_1(l,k)\check{\digamma}_{N}(\Xi_p^1(l,k))\nonumber\\
                    &\times K_1^\mathrm{T}(l,k)+\sum_{\nu=1}^{r_1}\check{\beta}_{1,\nu}H_{1,\nu}\check{\digamma}_{N}(\Xi_p^1(l,k))H^\mathrm{T}_{1,\nu}\label{t1-1b}
				\end{align}
                \end{small}
			\end{subequations}
		
			\begin{subequations}\label{t2-1}
				$\mathbf{Step\  2}$: For $(l,k)\in \mathcal{S}_1$
				\begin{small}
                \begin{align}
                    K_2(l,k)=&(1+\check{b}_3+\check{b}_4)\Xi_p^2(l,k)C_{N-1}^\mathrm{T}(l,k)\big(\check{\digamma}_{N-1}(\Xi_p^2(l,k))\big)^{-1}\label{t2-1a}\\
                    \Xi_u^2(l,k)=&(1+\check{b}_3+\check{b}_4)\Xi_p^2(l,k)-K_2(l,k)\check{\digamma}_{N-1}(\Xi_p^2(l,k))\nonumber\\
                    &\times K_2^\mathrm{T}(l,k)+\sum_{\nu=1}^{r_2}\check{\beta}_{2,\nu}H_{2,\nu}\check{\digamma}_{N-1}(\Xi_p^2(l,k))H^\mathrm{T}_{2,\nu}\label{t2-1b}
				\end{align}
                \end{small}
			\end{subequations}
			$\qquad\qquad\qquad\qquad\qquad \vdots$
			
			\begin{subequations}\label{t3-1}
				$\mathbf{Step\  N+1}$: For $(l,k)\in \mathcal{S}_{N}$
                \begin{small}
                \begin{align}
					K_{N+1}(l,k)=&(1+\check{b}_3+\check{b}_4)\Xi_p^{N+1}(l,k)\nonumber\\
					&\times C_{0}^\mathrm{T}(l,k)\big(\check{\digamma}_{0}(\Xi_p^{N+1}(l,k))\big)^{-1}\label{t3-1a}\\
					\Xi_u^{N+1}(l,k)=&(1+\check{b}_3+\check{b}_4)\Xi_p^{N+1}(l,k)-K_{N+1}(l,k)\nonumber\\
                    &\times \check{\digamma}_{0}(\Xi_p^{N+1}(l,k))K_{N+1}^\mathrm{T}(l,k)+\sum_{\nu=1}^{r_{N+1}}\check{\beta}_{{N+1},\nu}\nonumber\\
					&\times H_{{N+1},\nu}\check{\digamma}_{0}(\Xi_p^{N+1}(l,k))H^\mathrm{T}_{{N+1},\nu}\label{t3-1b}
				\end{align}
                \end{small}
			\end{subequations}
		\end{theorem}
		\begin{IEEEproof}
            Since the proof of each step is similar, for simplicity, only the proof of $\mathbf{Step\ 1}$ is given here. By using the square completion method for \eqref{cw4:b3b}, it can be obtained that
			\begin{align}\label{37}
				\Xi_u^1(l,k)=&\big(K_1(l,k)-K_1^*(l,k)\big)\check{\digamma}_{N}(\Xi_p^1(l,k))\nonumber\\
				&\times \big(K_1(l,k)-K_1^*(l,k)\big)^\mathrm{T}+(1+\check{b}_3+\check{b}_4)\Xi_p^1(l,k)\nonumber\\
				&-K_1^*(l,k)\check{\digamma}_{N}(\Xi_p^1(l,k))(K_1^*(l,k))^\mathrm{T}\nonumber\\
				&+\sum_{\nu=1}^{r_1}\check{\beta}_{1,\nu}H_{1,\nu}\check{\digamma}_{N}(\Xi_p^1(l,k))H_{1,\nu}^\mathrm{T}
			\end{align}
            where $K_1^*(l,k)=(1+\check{b}_3+\check{b}_4)\Xi_p^1(l,k)C_{N}^\mathrm{T}(l,k) \big(\check{\digamma}_{N}(\Xi_p^1(l,\\k))\big)^{-1}$ is minimized if and only if $K_1(l,k)=K_1^*(l,k)$, so the assert \eqref{t1-1a} is valid. Inserting \eqref{t1-1a} into \eqref{37}, one has
			\begin{align*}
                \Xi_u^1(l,k)=&(1+\check{b}_3+\check{b}_4)\Xi_p^1(l,k)-K_1(l,k)\check{\digamma}_{N}(\Xi_p^1(l,k))\nonumber\\
                &\times K_1^\mathrm{T}(l,k)+\sum_{\nu=1}^{r_1}\check{\beta}_{1,\nu}H_{1,\nu}\check{\digamma}_{N}(\Xi_p^1(l,k))H^\mathrm{T}_{1,\nu}
			\end{align*}
			This completes the proof.
		\end{IEEEproof}
        \begin{remark}
            So far, the event-triggered filtering problem of 2-D systems with asynchronous-delay under binary encoding-decoding schemes have been solved. Due to the existence of asynchronous-delay, the measurements are divided into $N$+1 regions for reconstruction, and the corresponding $N$+1-step recursive filter is given. Considering the impact of triggering error and encoding truncation error on the filter, the local upper bounds of prediction error covariance and filtering error covariance are given. The estimation gain parameters are obtained by minimizing the local upper bound of filtering error covariance. It is obvious from \eqref{t1-1}-\eqref{t3-1} that all information about the system model (triggering parameter, encoding truncation error, and relevant decoding information) are reflected in the filter design and has a direct impact on filtering performance.
		\end{remark}
	\subsection{Monotonicity of triggering parameters}
        As shown in Theorem~\ref{the1}, the evolution of the minimum upper bound of prediction error covariance and filtering error covariance depends on the triggering parameters, which is essentially reflected in the impact of the trigger error on the minimum upper bound of the estimated error covariance. Therefore, in this subsection, strict mathematical derivation will be used to discuss the relationship between the triggering error $\delta_{\tau}(l,k)$ and the local minimum upper bound of filtering error covariance $\Xi_u^{N+1-\tau}(l,k)$.
		
        For the sake of simplicity, only discuss $(l,k)\in \mathcal{S}_0$. The notations $\Xi_p^{1\delta_{N}}(l,k), \Xi_u^{1\delta_{N}}(l,k),$ and $K_1^{\delta_{N}}$ represent the local upper bound of error covariance and the estimation gain parameter with the triggering error $\delta_{N}(l,k)$. The following defines the operators $\mathcal{G}((l,k),\delta(l,k),K(l,k),\Xi(l,k))$ and $\mathcal{F}((l,k),\delta(l,k),K(l,k),\Xi(l,k))$,
		\begin{align}
			&\mathcal{G}((l,k),\delta(l,k),K(l,k),\Xi(l,k))\nonumber\\
			&\quad \triangleq(1+\check{b}_3+\check{b}_4)\big[I-K(l,k)C_{N}(l,k)\big]\nonumber\\
			&\quad\quad\times \Xi(l,k)\big[I-K(l,k)C_{N}(l,k)\big]^\mathrm{T}\nonumber\\
            &\quad\quad +\sum_{\nu=1}^{r_1}\check{\beta}_{1,\nu}H_{1,\nu}\check{\digamma}_{N}(\Xi(l,k))H_{1,\nu}^\mathrm{T}\nonumber\\
			&\qquad +K(l,k)\check{\Theta}(\bar{X}(l,k))K^\mathrm{T}(l,k)\label{38}\\
			&\mathcal{F}((l,k),\delta(l,k),K(l,k),\Xi(l,k))\nonumber\\
			&\quad \triangleq(1+\check{b}_1)\Xi_1(l,k-1)+(1+\check{b}_1^{-1})\Xi_2(l-1,k)\nonumber\\
			&\qquad +Q_1(l,k-1)+Q_2(l-1,k)\label{39}
		\end{align}
        where $\Xi_{\iota}(l,k)=(1+\check{b}_2)a^2_{\iota}(l,k)\text{tr}\{\Xi(l,k)\}I+(1+\check{b}_2^{-1})A_{\iota}(l,\\ k)\Xi(l,k)A_{\iota}^\mathrm{T}(l,k)$. Obviously, it can be seen that $\Xi_u^{1\delta_{N}}(l,k)=\\ \mathcal{G}((l,k),\delta_{N}(l,k),K_1^{\delta_{N}}(l,k),\Xi_p^{1\delta_{N}}(l,k))$.
		
        From the definition of $\check{\digamma}_{N}(\Xi_p^{1\delta_{N}}(l,k))$, it is easy to see that $\check{\digamma}_{N}(\Xi_p^{1\delta_{N}}(l,k))$ is monotonic about $\delta_{N}(l,k)$. Then, $\mathcal{G}((l,k),\\ \delta_{N}(l,k),K_1^{\delta_{N}}(l,k),\Xi_p^{1\delta_{N}}(l,k))$ is monotonic about $\delta_{N}(l,k)$ and $\Xi_p^{1\delta_{N}}(l,k)$.
		\begin{theorem}\label{the2}
            For given scalars $\sigma_{(s)}^{\gamma,1}, \sigma_{(s)}^{\gamma,1}, \rho_{(s)}^{\gamma,1},\rho_{(s)}^{\gamma,2},\varsigma_{s}^1,$ and $\varsigma_{s}^2$, if the conditions $\sigma_{(s)}^{\gamma,1}\leq\sigma_{(s)}^{\gamma,1},\rho_{(s)}^{\gamma,1}\geq\rho_{(s)}^{\gamma,2}$, and $\varsigma_{s}^1\leq \varsigma_{s}^2$, i.e. $0<\delta_{N}^1(l,k)\leq\delta_{N}^2(l,k)$, then the following inequality holds
			\begin{align*}
				\Xi_u^{1\delta_{N}^1}(l,k)\leq\Xi_u^{1\delta_{N}^2}(l,k)
			\end{align*}
			for every $k\in[0,i_0]$, $k\in[0,j_0]$.
		\end{theorem}
        \begin{IEEEproof}
            In view of Theorem 1, $\Xi_u^{1\delta_{N}}(l,k)$ can be minimized if and only if $K_1^{\delta_{N}}(l,k)=K_1^{*{\delta_{N}}}(l,k)=(1+\check{b}_3+\check{b}_4)\Xi_p^{1\delta_{N}}(l,k)C_{N}^\mathrm{T}(l,k)(\check{\digamma}(\Xi_p^{1\delta_{N}}(l,k)))^{-1}$, so
			\begin{align*}
				&\mathcal{G}((l,k),\delta_{N}(l,k),K_1^{*{\delta_{N}}}(l,k),\Xi_p^{1\delta_{N}}(l,k))\\
				&\qquad\qquad \leq \mathcal{G}((l,k),\delta_{N}(l,k),K_1^{\delta_{N}}(l,k),\Xi_p^{1\delta_{N}}(l,k))
			\end{align*}
            Next, by using mathematical induction to prove this theorem.
            ${\it Initial\ Step:}$According to the initial condition $\Xi_u^{1\delta_{N}}(l,0)$ and $\Xi_u^{1\delta_{N}}(0,k)$ are independent of the triggering error $\delta_{N}(l,k)$, so $\Xi_p^{1\delta_{N}^1}(1,1)=\Xi_p^{1\delta_{N}^2}(1,1)$, due to \eqref{38}, it is easy to know that
			\begin{align*}
				\Xi_u^{1,\delta_{N}^1}(1,1) \leq \Xi_u^{1,\delta_{N}^2}(1,1).
			\end{align*}
            ${\it Inductive\ Step:}$ Assume $\Xi_u^{1\delta_{N}^1}(l,k)\leq \Xi_u^{1\delta_{N}^2}(l,k)$ is valid for $(l,k)\in\{(l_0,k_0)|l_0\in[1,i_0],k_0\in[1,j_0]; l_0+k_0=\flat\}$ with $\flat\in[2,i_0+j_0-1]$, then, for $(l,k)\in\{(l_0,k_0)|l_0\in[1,i_0],k_0\in[1,j_0];l_0+k_0=\flat+1\}$, one has
			 \begin{align*}
                \Xi_u^{1,\delta_{N}^1}(l,k)&=\mathcal{G}((l,k),\delta_{N}^1(l,k),K_1^{*\delta_{N}^1}(l,k),\Xi_p^{1\delta_{N}^1}(l,k))\\
			 	&\leq \mathcal{G}((l,k),\delta_{N}^1(l,k),K_1^{*\delta_{N}^2}(l,k),\Xi_p^{1\delta_{N}^1}(l,k))\\
			 	&\leq \mathcal{G}((l,k),\delta_{N}^2(l,k),K_1^{*\delta_{N}^2}(l,k),\Xi_p^{1\delta_{N}^2}(l,k))\\
			 	&=\Xi_u^{1,\delta_{N}^2}(l,k).
			 \end{align*}
		 	This completes the proof.
		\end{IEEEproof}
		\begin{remark}
            It is noteworthy that, according to Theorem~\ref{the2}, the upper bound of the filtering error covariance does not decrease as the triggering error increases. Intuitively speaking, a larger triggering error leads to a higher frequency of update triggers, resulting in increased utilization of network resources by each channel, thereby exacerbating the tension on network resource utilization. At the same time, an increase in triggering error implies that more signals will be transmitted for estimation, which to a certain extent enhances the accuracy of the estimation. Therefore, setting the triggering parameters reasonably can achieve a satisfactory compromise between energy saving and estimation performance.
		\end{remark}

\section{Conclusion}
    To summarize, this paper has contributed some of the first efforts to the problem of recursive filtering with asynchronous-delay. First, an ETMs have been introduced under the premise of asynchronous-delay to reduce the number of signal transmissions, improve communication quality, and enable measurement reconstruction after triggering. Then, the mathematical quantification of signal distortion during communication under binary EDSs has been given. Then, through mathematical induction, a local upper bound on the estimation error covariance has been found, and the estimation gain parameters have been given by minimizing the upper bound on the estimation error covariance. In addition, the monotonicity of the filtering performance with respect to the triggering parameters has been further discussed. Future research topics include extending existing filter design results to have more complex dynamics and communication mechanisms, such as dynamic ETMs with random communication delay, and unknown input estimation problems with asynchronous-delay \cite{Wang_TSIP22,Zhu_ATO22,Li_TNNLS20,Ge_TCB19,Cui_AMC19,Song_IT19}.

\end{document}